\documentclass[twocolumn,showpacs,amsmath,prb,superscriptaddress]{revtex4-2}
\usepackage{epsfig,psfrag}
\usepackage{amsbsy}
\usepackage{amsfonts}
\usepackage{bbm}
\usepackage{tabularx}
\usepackage{euscript}
\usepackage{amsfonts}
\usepackage{exscale}
 \usepackage{slashed}
\usepackage{comment}

\usepackage{amsmath,amssymb}
\usepackage{graphicx}
\usepackage{dcolumn}
\usepackage{bm}
\usepackage{braket}
\usepackage{verbatim}
\usepackage{float}
\usepackage{bbold}
\usepackage{color}
\usepackage{xcolor}
\usepackage{relsize}
\usepackage{amsthm}
\usepackage{enumerate}
\usepackage{soul,xcolor}
\usepackage[normalem]{ulem}
\usepackage[T1]{fontenc}
\usepackage[colorlinks=true ,urlcolor=blue,urlbordercolor={0 1 1}]{hyperref}

\newcommand{\norm}[1]{\mathopen{:}#1\mathclose{:}}

\newcommand{\bfk}{\mathbf{k}}
\newcommand{\bfK}{\mathbf{K}}

\newcommand{\bfQ}{\mathbf{Q}}

\newcommand{\ph}{\mathrm{ph}}
\newcommand{\pp}{\mathrm{pp}}

\newcommand{\CDW}{\mathrm{CDW}}
\newcommand{\PDW}{\mathrm{PDW}}
\newcommand{\VHS}{\mathrm{VHS}}
\newcommand{\MF}{\mathrm{MF}}

\begin{document}
\graphicspath{{figures/}}

\title{
Emergent Attraction and Competing Orders near the Van Hove Singularity in Rhombohedral $n$-layer Graphene
}

\date{\today}

\author{Jing-Yu Zhao}
\affiliation{ Condensed Matter Theory Center and Joint Quantum Institute, Department of Physics, University of Maryland, College Park, Maryland 20742, USA}
\author{Yang-Zhi Chou}
\affiliation{ Condensed Matter Theory Center and Joint Quantum Institute, Department of Physics, University of Maryland, College Park, Maryland 20742, USA}
\author{Sankar Das Sarma}
\affiliation{ Condensed Matter Theory Center and Joint Quantum Institute, Department of Physics, University of Maryland, College Park, Maryland 20742, USA}

\begin{abstract}
Unconventional superconductivity in rhombohedral $n$-layer graphene exhibits a rich phase diagram with multiple distinct superconducting regions, whose microscopic origins remain unclear. 
Here, we investigate the competition between charge-density-wave (CDW) and pair-density-wave (PDW) instabilities near the inner van Hove singularities (VHSs), using a patch renormalization-group approach with the realistic band structures incorporated.
We identify two key findings.
First, we find an effective attractive interaction emerging from purely repulsive interactions through RG flow in the presence of $C_3$ symmetry breaking.
The resulting interaction structure, with two repulsive couplings and one attractive coupling, gives rise to competing CDW and PDW instabilities.
Second, the competition is controlled by the detuning from the VHS. 
While CDW dominates near the VHS filling and at higher temperatures, PDW becomes the leading instability on either side of the VHS and at lower temperatures.
This mechanism naturally produces two superconducting regions separated by a CDW-dominated regime, providing a possible qualitative explanation for the experimentally observed separation between SC1 and SC2.
\end{abstract}
\maketitle{}

\textit{Introduction}---
The recent discovery of superconductivity (SC) in rhombohedral $n$-layer graphene (R$n$G) 
\cite{HanRnGSC2025b,JIAli2025,huaR6G2026, zhengCSC2026,kalantreFermiology2026} has attracted considerable attention. 
Of particular interest is the possibility that superconductivity emerges from a spin- and valley-polarized normal state with broken time-reversal symmetry.
Despite extensive theoretical efforts
\cite{geierCSC2025,chouR4GSC2025,yangPDW2025,parraRPA2025,gaggioliVAV2025,youKL2022,qinFRG2023,chouQM2026,christosPDW2025,sauAnomalousHall2024,zhuCSC2026,gilPDW2026,dongQM2026,qinR4G2026,chenDiode2025,jahinKohnLuttinger2026,liBerry2025,maymannBerry2026,patriRnG2025,wangCSC2026,yoonQMl2026a, herasymchuk2026, primeauGeometry2026,
wangWignerSC2026,desrochersWC2026, dongWC2026b, kimWC2026, kimWC2026a, miaoWignerSC2026, abouelkomsanCrystals2026, zhangWCSC2026},
the microscopic pairing mechanism and its relation to the observed phase diagram remain under debate.
While chiral superconductivity driven by screened Coulomb interactions has been predicted within Kohn-Luttinger based approaches \cite{geierCSC2025,yangPDW2025,chouR4GSC2025,jahinKohnLuttinger2026,parraRPA2025,qinR4G2026,dongQM2026,maymannBerry2026}, 
the recent observation of Wigner crystallization in R4G \cite{hanWigner2026} has motivated alternative proposals linking superconductivity to Wigner crystalline correlations \cite{wangWignerSC2026,desrochersWC2026,dongWC2026b,kimWC2026,kimWC2026a,miaoWignerSC2026,abouelkomsanCrystals2026, zhangWCSC2026}.
Nevertheless, a comprehensive theoretical understanding of the observed phase diagram remains lacking.

One particular puzzle is the occurrence of multiple nearby superconducting regions.
Experimentally, at least two distinct superconducting regions, SC1 and SC2, have been reported on opposite sides of a line of enhanced resistivity
\cite{HanRnGSC2025b,huaR6G2026,JIAli2025, zhengCSC2026}. 
Their separation may reflect competition with other electronic orders, a change in the parent metallic state, or a combination of both.
The former possibility is further suggested by observations of nearby Wigner-crystalline \cite{hanWigner2026} and nematic phases \cite{JIAli2025}, while recent quantum-oscillation measurements \cite{zhengCSC2026} suggest a Lifshitz transition between the two superconducting regions.

These observations motivate a study of competing electronic orders near the van Hove singularities (VHSs) associated with the Lifshitz transition.
A patch renormalization group (RG) approach of VHSs \cite{honerkampRG2001,nandkishoreVHS2012, gilPDW2026, leeHVHS2024} captures the mutual renormalization of particle-hole (PH) and particle-particle (PP) channels on equal footing, 
making it particularly well suited for studying competing electronic orders and resultant quantum phases.
Recently, Ref.~\cite{gilPDW2026} investigated the competition between charge-density-wave (CDW) and pair-density-wave (PDW) instabilities for different VHS geometries.
However, it remains to be understood how the results are related to the realistic experimental phase diagram with finite detuning from the VHS energy. 
And more importantly, as we discuss later, the possibility of $C_3$ symmetry breaking has not been fully explored.

\begin{figure}[t]
    \centering
    \includegraphics[width=0.95\linewidth]{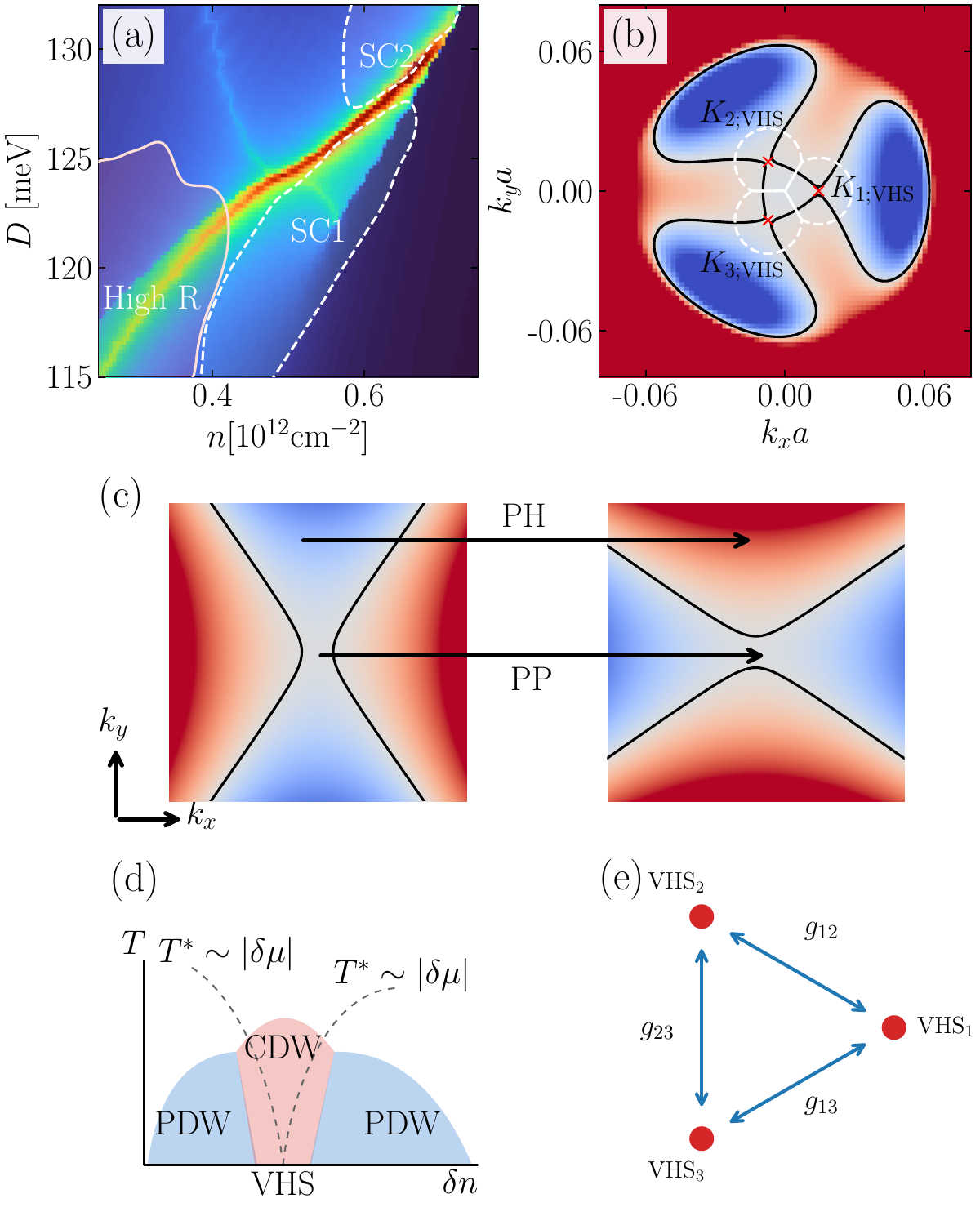}
    \caption{
    (a) Density of states of R$4$G as a function of carrier density $n$ and displacement field $D$. 
    The shaded regions indicate the approximate locations of the experimentally observed SC1 and SC2 regions, extracted from Ref.~\cite{HanRnGSC2025b} and subjected to a linear transformation of the displacement field. 
    (b) Representative Fermi surface near the inner VHSs at $D=120$ meV and $n=0.36\times 10^{12}\mathrm{cm}^{-2}$ of R4G. 
    Red and blue indicate unoccupied and occupied states, respectively.
    The VHS momenta $\mathbf{K}_{a;\VHS}$ are marked by red crosses.
    The dashed white circles with radius $k_{\mathrm{cut}}a=0.015$ indicate the patches centered at the three inner VHSs used in our calculations, with $a$ the graphene lattice constant.
    (c) Schematic illustration of PH and PP fluctuations between two quadratic VHSs rotated by $90^\circ$ relative to each other, with a finite chemical potential detuning.
    (d) Schematic phase diagram illustrating the competition between CDW and PDW instabilities. 
    CDW dominates near VHS filling, whereas PDW is favored on either side. 
    The dashed curves indicate the crossover scale $T\sim|\delta\mu|$, separating the VHS-dominated and low-energy Fermi-surface regimes.
    (e) Three-patch interaction model involving the three inner VHSs, with interpatch couplings $g_{12}$, $g_{13}$, and $g_{23}$.
    }
    \label{fig:illu}
\end{figure}

In this work, we investigate interaction-driven instabilities near an inner VHS line that may separate the experimentally observed SC1 and SC2 regions, as illustrated in Fig.~\ref{fig:illu}.
We employ a weak-coupling patch RG approach based on realistic band structures of R$n$G, allowing for a finite detuning $\delta\mu$ of the chemical potential from the VHS energy.
Given the strongly correlated nature of R$n$G, our patch RG analysis is not intended to provide quantitative estimates of the transition temperatures $T_c$, but rather to reveal the qualitative mechanisms and phase structure of the competing instabilities.

Our theory reveals two key findings.
First, direct PH fluctuations amplify an infinitesimal $C_3$ asymmetry in the interpatch interactions, driving initially repulsive couplings toward a strong-coupling trajectory characterized by two repulsive couplings and one attractive coupling, denoted by $(+,+,-)$.
Second, this emergent interaction structure gives rise to competing nematic CDW and PDW instabilities, 
whose relative strengths depend sensitively on the detuning from the VHS.
While CDW order dominates near the VHS filling, PDW superconductivity becomes the leading instability upon detuning away from the VHS.

This mechanism naturally gives rise to PDW superconducting regions on both sides of the VHS line, separated by a charge-ordered regime along the VHS line.
Our results thus provide a possible qualitative explanation for the experimentally observed separation of superconducting regions and their emergence slightly away from the maximal resistivity line. 
We also examine the temperature evolution of the competing orders in a toy model and identify that the charge-ordered regime may be related to the experimentally observed ``undetermined metal'' \cite{HanRnGSC2025b}.

\textit{VHSs of R$n$G} ---
In Fig.~\ref{fig:illu}(a), we show the density of states (DOS) as a function of carrier density $n$ and displacement field $D$ for spin- and valley- polarized R4G. 
An X-shaped DOS peak emerges from two distinct sets of VHSs.
Similar structures are also found in R5G and R6G. 
The branch extending from the lower left to the upper right corresponds to the touching of the inner and outer Fermi pockets, whereas the other branch is associated with the evolution of an annular Fermi surface into three outer pockets. 
Notably, the experimentally observed superconducting regions, indicated by the shaded areas extracted from Ref.~\cite{HanRnGSC2025b}, 
exhibit a qualitative correspondence with the inner VHS line.
Motivated by this observation, we focus on the inner VHSs. 

A representative Fermi surface with three inner VHS points connecting the inner and outer pockets is shown in Fig.~\ref{fig:illu}(b). 
Their momenta are denoted by $\mathbf{K}_{a;\VHS}$ ($a=1,2,3$), related by $120^\circ$ rotations. 
As $D$ increases, the VHS momenta move toward the valley center and eventually merge at $D\simeq132$ meV.
We focus on $D=120$ meV throughout the following analysis, while the qualitative behaviors are similar at other displacement fields.

Before a detailed RG analysis, a simple observation already suggests qualitatively different physics at and away from VHS filling. 
We consider both PH scattering $\bfQ_\CDW=\bfK_{b;\VHS}-\bfK_{a;\VHS}$ 
and PP scattering $\bfQ_\PDW=\bfK_{a;\VHS}+\bfK_{b;\VHS}$ 
between patch $a$ and $b$. 
For illustration, we show two quadratic VHSs with identical dispersions but rotated by $90^\circ$ relative to each other in Fig.~\ref{fig:illu}(c). 
The chemical potential is slightly detuned from the VHS energy by $\delta\mu$.
At energy scales $E\gtrsim|\delta\mu|$, interpatch scattering predominantly connects occupied and unoccupied states, 
as illustrated by the PH process in Fig.~\ref{fig:illu}(c), enhancing PH fluctuations in the CDW channel.
In contrast, at lower energy scales $E\lesssim|\delta\mu|$, interpatch scattering near the VHSs tends to connect states with the same occupation, as indicated by the PP process in Fig.~\ref{fig:illu} (c). 
The enhanced DOS near the VHSs therefore favors PP fluctuations in the PDW channel. 

In R$n$G, the VHSs are not strictly quadratic and are related by $120^\circ$ rather than $90^\circ$ rotations,
but the physics is similar, as we discuss later. 
This picture leads to the schematic phase diagram illustrated in Fig.~\ref{fig:illu}(d).
At VHS filling, PH fluctuations dominate down to the lowest energy scales, favoring a nematic CDW state.
Upon slight detuning to either side of the VHS, PH fluctuations remain important at high energies $E\gtrsim |\delta \mu|$, while PDW becomes the leading instability at lower energies.

\textit{RG of three VHSs}--- 
The above picture leaves open how PDW superconductivity can emerge from purely repulsive interactions.
We show that an attractive channel can dynamically emerge from RG flow once a small $C_3$ asymmetry is considered. 
We construct a low-energy patch model for the single-flavor fermions near the three VHSs as illustrated in Fig.~\ref{fig:illu}(e): 
\begin{equation}\label{eqn:patchH}
    H=\sum_a\xi_a(\mathbf{k})\psi_a^\dagger\psi_a
    +\sum_{a<b}g_{ab}\psi_a^\dagger\psi_b^\dagger\psi_b\psi_a~,
\end{equation}
where momentum labels and integrations over the patches are implicit. 
$\xi_a(\mathbf{k})=\epsilon(\bfk+\mathbf{K}_{a;\VHS})-\mu$ denotes the dispersion near each saddle point, with $\epsilon(\bfk)$ from the realistic band structure.

Following Refs.~\cite{honerkampTfRG2001,qinFRG2023}, we employ the temperature-flow RG, which treats PH and PP fluctuations on equal footing 
\footnote{
In the conventional momentum-shell RG treatment \cite{shankarRG1994}, the direct PH contribution is absent from the one-loop flow, whereas in temperature-flow RG it exhibits a single-logarithmic enhancement near a VHS
}. 
Unlike conventional VHS problems with double-logarithmic divergences, the patches in R$n$G are generally not nested and exhibit single-logarithmic divergences associated with the DOS divergence. 
We use $\ell=\ln(E_{\mathrm{cut}}/T)$ as the RG flow parameter. 
At one-loop order, the interpatch couplings obey 
\begin{equation}\label{eqn:beta}
    \frac{dg_{ab}}{d\ell} =\frac{m}{2\pi^2}
    \left[
    \left(\alpha_{\rm ph}^e+\alpha_{\rm pp}\right)g_{ab}^2
    -\alpha_{\rm ph}^d g_{ac}g_{bc}
    \right]~.
\end{equation}
Here, $c\neq a,b$ denotes the remaining patch, 
and $m$ is the effective mass obtained from the quadratic expansion of the VHS dispersion, $\xi_{a}(\bfk)\sim k_{a;\parallel}^2/(2\sqrt{\eta}m)-\sqrt{\eta}k_{a;\perp}^2/(2m)$. 
$\alpha_\ph^d$ denotes the direct PH contribution, which equals unity for a quadratic saddle point at VHS filling. 
We further define $\alpha=\alpha_{\mathrm{ph}}^e+\alpha_{\mathrm{pp}}$, which characterizes the competition between the exchange PH ($\alpha_{\mathrm{ph}}^e>0$) and PP ($\alpha_{\mathrm{pp}}<0$) contributions 
\footnote{
We note that our results differ from those of Ref.~\cite{gilPDW2026} in three respects: the absence of the $\alpha_{\ph}^d$ term, the opposite sign of $\alpha_{\ph}^e$, and a different coefficient for $\alpha_{\pp}$.
We have carefully checked our expressions using both temperature-flow RG and the frequency-shell RG scheme employed in Ref.~\cite{gilPDW2026}, obtaining consistent results}. 
Details of the derivation are provided in the End Matter and Supplemental Material \cite{SM}.

We first consider three quadratic VHSs at van Hove filling, for which $\alpha_{\ph}^d=1$ and $\alpha$ is independent of the RG scale $\ell$.
Along the $C_3$-symmetric trajectory with $g_{12}=g_{13}=g_{23}\equiv g$: 
\begin{equation}\label{eqn:flow_sym}
    g(\ell)=\frac{g_0}{1+\frac{m}{2\pi^2}(1-\alpha)g_0\ell}~.
\end{equation}
For realistic band parameters, we typically find $-1<\alpha<1$, such that an initially repulsive coupling $g_0>0$ is marginally irrelevant and flows to zero. 
This behavior reflects the familiar screening of repulsive interactions by PH fluctuations, where the direct PH contribution dominates over the exchange PH and PP contributions.

\begin{figure}[t]
    \centering
    \includegraphics[width=0.99\linewidth]{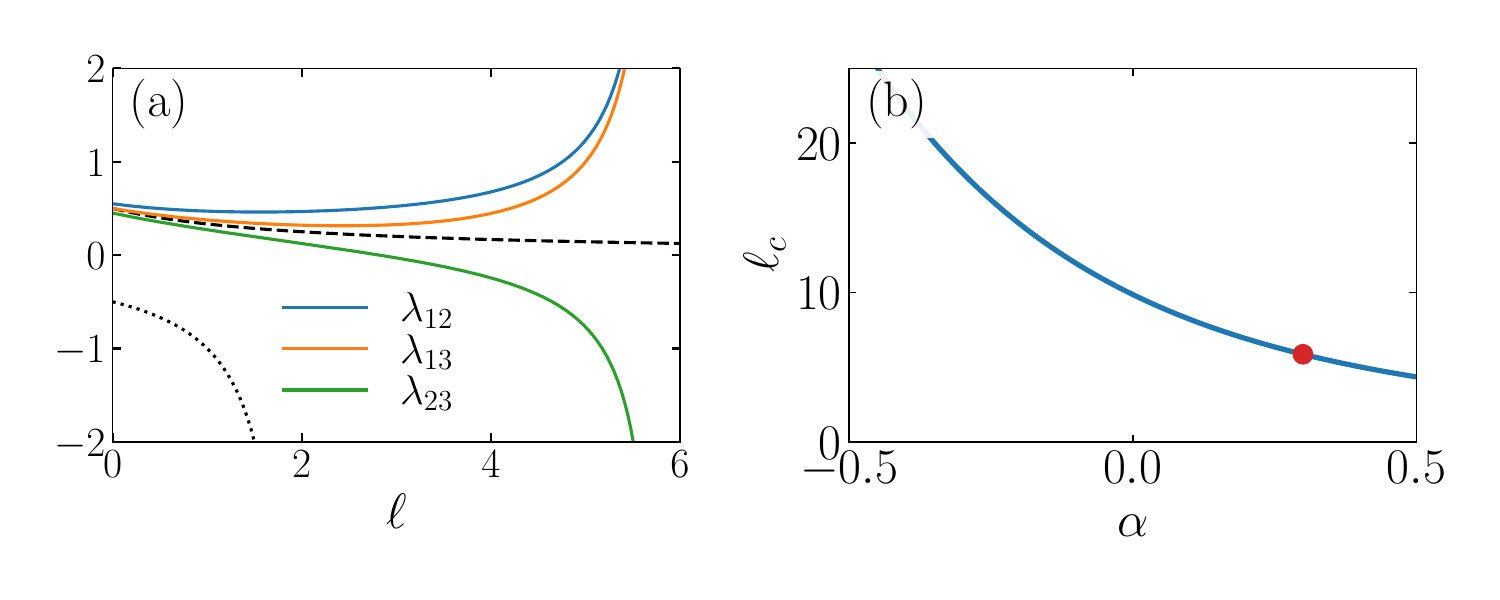}
    \caption{RG flows of the interpatch interactions and their divergence scale.
    We use the dimensionless couplings $\lambda_{ab}=mg_{ab}/(2\pi^2)$ and $\alpha=\alpha_{\mathrm{ph}}^e+\alpha_{\mathrm{pp}}$, $\alpha_{\ph}^d=1$.
    (a) RG flows for weakly $C_3$-asymmetric repulsive interactions, with initial values $(\lambda_{12},\lambda_{13},\lambda_{23})=(0.55,0.50,0.45)$ and $\alpha=0.3$. 
    The couplings flow toward the strong-coupling $(+,+,-)$ structure. 
    The black dashed and dotted curves show the $C_3$-symmetric trajectories with repulsive and attractive initial interactions with $\lambda=\pm 0.5$, respectively.
    (b) Divergence scale $\ell_c$ as a function of $\alpha$, using the same asymmetric initial interactions as in (a).
    The red dot marks $\alpha=0.3$, corresponding to the flow shown in (a).
    }
    \label{fig:flows}
\end{figure}

However, the $C_3$-symmetric RG trajectory is unstable against $C_3$-breaking perturbations. 
Consider an infinitesimal perturbation $g_{ab}=g(\ell)(1+\delta \lambda_{ab})$.  
To the linear order of Eq.~\eqref{eqn:beta}, we obtain the symmetric mode $\delta \lambda_s(\ell)=\delta \lambda_s(0)f(\ell)^{-1}$ and 
the doubly degenerate $C_3$-breaking mode $\delta \lambda_n(\ell)=\delta \lambda_n(0)f(\ell)^{(\alpha+2)/(1-\alpha)}$, 
where $f(\ell)=1+\frac{m}{2\pi^2}(1-\alpha)g_0\ell$. 
The latter grows to infinity as $\ell \to \infty$ for $-1<\alpha<1$. 

The instability ultimately drives the RG flow toward a stable strong-coupling $(+,+,-)$ ray with two repulsive couplings and one attractive coupling:
\begin{equation}
    g_{12}(\ell)=g_{13}(\ell)=\frac{(1+\alpha)g_c}{\ell_c-\ell},\qquad
    g_{23}(\ell)=-\frac{g_c}{\ell_c-\ell},
\end{equation}
where $g_c=2\pi^2/[m(\alpha^2+\alpha+1)]$ and $\ell_c$ is the divergence scale set by the initial conditions. 

Representative RG flows for $C_3$-symmetric and symmetry-breaking initial interactions are shown in Fig.~\ref{fig:flows}. 
Along the symmetric trajectory, repulsive interactions flow to zero, whereas attractive interactions diverge toward strong coupling. 
In contrast, a weak initial $C_3$ asymmetry drives purely repulsive interactions toward the strong-coupling $(+,+,-)$ structure. 
The structure is robust against the choice of initial interactions.
The divergence scale $\ell_c$ depends on both the initial couplings and $\alpha$. 
As shown in Fig.~\ref{fig:flows}(b), $\ell_c$ is substantially larger for $\alpha<0$ than for $\alpha>0$. 

We therefore obtain an effective attractive interaction from purely initial repulsive couplings. 
This is reminiscent of the Kohn-Luttinger mechanism \cite{kohn1965}, in which PH fluctuations can generate an effective attraction at higher angular momentum channels. 
Once $g_{23}$ becomes attractive, the PP fluctuation can further enhance the attraction and favor PDW pairing. 
Unlike the conventional Fermi-surface case, however, the PDW instability near VHS filling competes with a similarly enhanced CDW instability.

\textit{Test vertices}---
To identify the leading instability, we introduce test vertices $\Phi_{ab}\psi_a^\dagger\psi_b$ and $\Delta_{ab}\psi_a\psi_b$ for the CDW and PDW channels, respectively.
Their one-loop RG equations are
\begin{equation}\label{eqn:test_pph}
    \frac{d\log\Phi_{ab}}{d\ell}
    =\frac{m}{2\pi^2}\alpha_{\mathrm{ph}}^eg_{ab},
    \quad
    \frac{d\log\Delta_{ab}}{d\ell}
    =\frac{m}{2\pi^2}\alpha_{\mathrm{pp}}g_{ab}.
\end{equation}
The leading instability is identified by comparing the growth rates of the test vertices as $\ell\to\ell_c$.
Along the $(+,+,-)$ trajectory, the dominant instability is determined by the relative magnitudes of $(1+\alpha)\alpha_{\mathrm{ph}}^e$ for CDW and $-\alpha_{\mathrm{pp}}$ for PDW.
Details of the susceptibility analysis are provided in the End Matter.

\begin{figure}[t]
    \centering
    \includegraphics[width=0.99\linewidth]{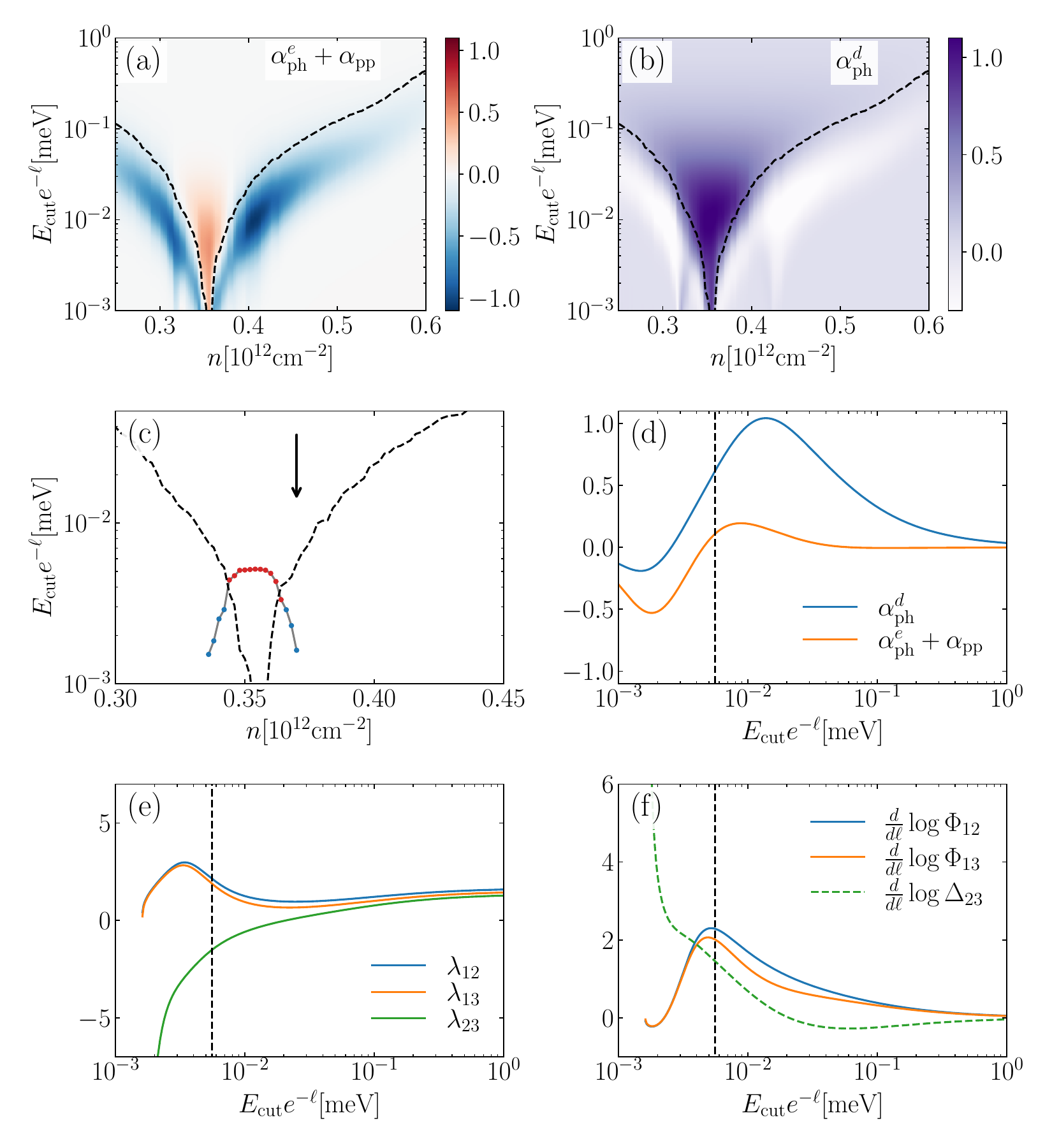}
    \caption{
    RG analysis using realistic band parameters for R4G, with $E_{\rm cut}=3$ meV as the initial RG cutoff.
    (a) $\alpha=\alpha_{\mathrm{ph}}^e+\alpha_{\rm pp}$ as a function of carrier density $n$ and RG scale $E_{\rm cut} e^{-\ell}$ at fixed displacement field $D=120$ meV.
    The dashed line indicates the detuning scale $|\mu-E_{\VHS}|$ from the VHS.
    (b) $\alpha_{\mathrm{ph}}^d$ as a function of $n$ and $E_{\rm cut} e^{-\ell}$ at the same displacement field $D=120$ meV. 
    (c) Strong-coupling scale $E_{\rm cut}e^{-\ell_c}$ for initial interactions $(\lambda_{12},\lambda_{13},\lambda_{23})=(1.65,1.50,1.35)$.
    Red and blue dots indicate CDW- and PDW-dominated instabilities, respectively, 
    determined by comparing the growth rates $\frac{d}{d\ell}\log\Phi_{12}(\ell_c)$ and $\frac{d}{d\ell}\log\Delta_{23}(\ell_c)$ near the RG divergence. 
    The black arrow marks the density $n=0.37\times10^{12}\mathrm{cm}^{-2}$ used in panels $(d-f)$. 
    (d) RG scale dependence of $\alpha=\alpha_{\mathrm{ph}}^e+\alpha_{\mathrm{pp}}$ and $\alpha_{\mathrm{ph}}^d$ at fixed density line cut $n=0.37\times10^{12}\mathrm{cm}^{-2}$.
    The dashed line marks the detuning scale $|\mu-E_{\VHS}|$.
    (e) RG flows of the interpatch couplings $\lambda_{ab}$ at the same density $n=0.37\times10^{12}\mathrm{cm}^{-2}$. 
    While $\lambda_{12}$ and $\lambda_{13}$ remain repulsive, $\lambda_{23}$ becomes attractive under flow.
    Due to the sign change of $\alpha$ shown in (d), $\lambda_{12}, \lambda_{13}$ finally decay to 0 and only $\lambda_{23}$ diverges.
    (f) Corresponding RG evolution of the test-vertex growth rates  $\frac{d}{d\ell}\log \Phi$ and $\frac{d}{d\ell}\log \Delta$. 
    Only the PDW test vertex $\Delta_{23}$ exhibits divergent growth, indicating a leading PDW superconducting instability.
    }
    \label{fig:rg_real}
\end{figure}

\begin{figure}[t]
    \centering
    \includegraphics[width=0.99\linewidth]{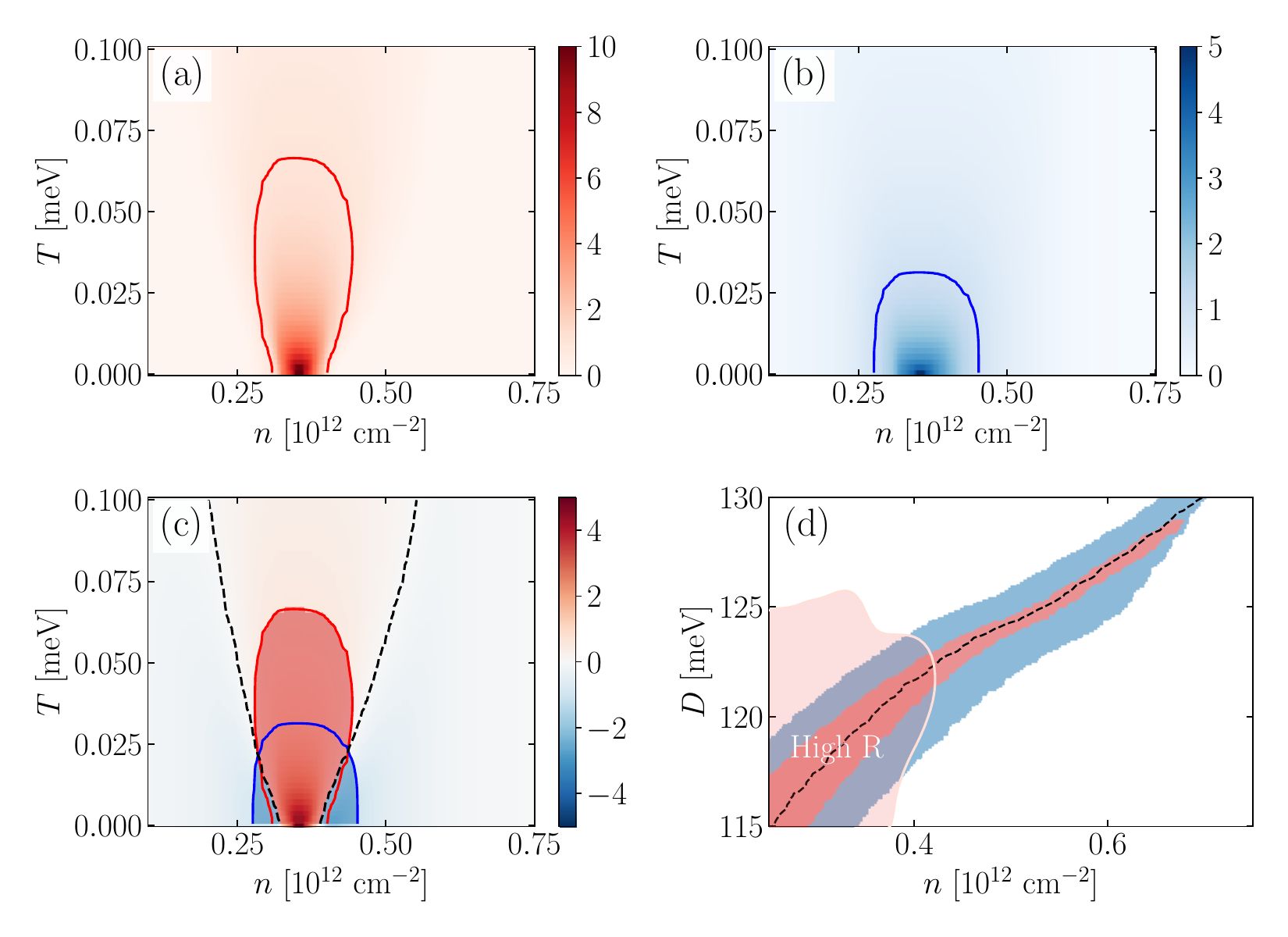}
    \caption{
    Instability susceptibilities for the toy model Eq.~\eqref{eqn:patch_int} 
    with $g=1.01\times10^3$ meV$\cdot$nm$^2$ and $v=0.51\times10^3$ meV$\cdot$nm$^2$, 
    as functions of carrier density $n$ and temperature $T$. 
    (a) CDW channel $-g\int_{\bf k}\Pi_{\mathrm{ph}}(\xi_{1;\bf k},\xi_{2;\bf k})$.
    (b) PDW channel $v\int_{\bf k}\Pi_{\mathrm{pp}}(\xi_{2;\bf k},\xi_{3;-\bf k})$.
    The red and blue curves indicate the respective mean-field instability boundaries, where the corresponding quantities reach unity. 
    (c) Overlay of the two instability regions. 
    The background color shows the difference between the dimensionless CDW and PDW kernels. The dashed line marks their equality.
    The regions where the CDW and PDW instabilities dominate are shaded in red and blue, respectively. 
    (d) Leading instability at zero temperature as a function of carrier density and displacement field, using the same coupling constants. 
    The red-shaded region indicates the experimentally observed high-resistivity regime \cite{HanRnGSC2025b}, where disorder is expected to suppress VHS-related physics.
    }
    \label{fig:sus}
\end{figure}

\textit{RG with realistic band structures. ---}
We now turn to realistic band structures of R4G 
\cite{kalantreFermiology2026,auerbachIsospin2025, SM}, 
allowing the chemical potential to be detuned from the VHS energy.
Figs.~\ref{fig:rg_real}(a,b) show the RG coefficients as functions of carrier density $n$ and RG scale $E_{\rm cut}e^{-\ell}$ for a fixed displacement field $D=120$ meV. 
Similar qualitative behaviors are found at other displacement fields. 

At VHS filling, the exchange PH contribution $\alpha_{\mathrm{ph}}^e>0$ dominates over the PP contribution $\alpha_{\mathrm{pp}}<0$, such that $\alpha>0$. 
Upon detuning from the VHS, the exchange PH contribution still dominates at high energy scales $E_{\rm cut}e^{-\ell}\gtrsim |\delta\mu|$, 
but gives way to the PP contribution at low energy scales $E_{\rm cut}e^{-\ell}\lesssim|\delta\mu|$, where $\alpha$ turns negative.
Similarly, the direct PH coefficient $\alpha_{\mathrm{ph}}^d$ approaches unity near VHS filling but is suppressed toward zero below the detuning scale.
These behaviors are consistent with the qualitative picture illustrated in Fig.~\ref{fig:illu}(c).

We further solve the RG flow equations using the realistic RG coefficients to determine the leading instabilities.
All flows start at $E_{\mathrm{cut}}=3$ meV, with initial couplings $\frac{m}{2\pi^2}(g_{12},g_{13},g_{23})=(1.65,1.50,1.35)$.
The resulting divergence scale and leading instabilities are shown in Fig.~\ref{fig:rg_real}(c), where red and blue dots indicate the leading CDW and PDW instabilities, respectively.
While the absolute divergence scales depend sensitively on the initial couplings, we focus on the qualitative structure of the phase diagram. 

A representative flow favoring PDW superconductivity is shown in Figs.~\ref{fig:rg_real}(d-f) at $n=0.37\times10^{12} \mathrm{cm}^{-2}$.
At high energies $E_{\rm cut}e^{-\ell}\gtrsim|\delta\mu|$, $\alpha>0$ and $\alpha_{\mathrm{ph}}^d\sim1$, 
such that strong PH fluctuations amplify the $C_3$ asymmetry and drive the initially repulsive interactions toward a $(+,+,-)$ structure.
At lower energies $E_{\rm cut}e^{-\ell}\lesssim|\delta\mu|$, $\alpha<0$ and $\alpha_{\mathrm{ph}}^d\sim0$.
Consequently, the repulsive couplings $g_{12}$ and $g_{13}$ are suppressed, while the attractive coupling $g_{23}$ diverges toward strong coupling.
This two-stage behavior is also reflected in the test vertices.
While the CDW vertices $\Phi_{12}$ and $\Phi_{13}$ are initially enhanced above the detuning scale, 
the PDW vertex $\Delta_{23}$ grows rapidly below this scale and eventually dominates.

\textit{Leading instabilities of a toy model}---
Motivated by the $(+,+,-)$ coupling structure emerging from RG, we consider a simplified patch model with interaction 
\begin{equation}\label{eqn:patch_int}
    H_{\rm int} = g\norm{\psi_1^\dagger\psi_2^\dagger\psi_2\psi_1}
    +g\norm{\psi_1^\dagger\psi_3^\dagger\psi_3\psi_1}
    -v\norm{\psi_2^\dagger\psi_3^\dagger\psi_3\psi_2}~.
\end{equation}
where $g,v>0$ are phenomenological parameters, and we focus on the qualitative structure of the phase diagram rather than its parameter dependence. 
We evaluate the CDW and PDW susceptibilities as functions of carrier density and temperature, with the corresponding instability conditions $-g\int_\bfk\Pi_{\mathrm{ph}}(\xi_{1;\bfk},\xi_{2;\bfk})=1$ and $v\int_\bfk\Pi_{\mathrm{pp}}(\xi_{2;\bfk},\xi_{3;-\bfk})=1$, respectively.
These criteria identify the leading instability of the symmetric Fermi surface, but do not necessarily determine the actual ordered ground state. 
Nevertheless, as shown by the mean-field calculations including charge redistribution in the Supplemental Material \cite{SM}, the qualitative tendency in Fig.~\ref{fig:sus} remains robust. 

As shown in Fig.~\ref{fig:sus}, the CDW instability extends to higher temperatures but is confined to a relatively narrow density range near the VHS.
In contrast, the PDW instability occurs at lower temperatures but spans a broader range of carrier densities.
Comparing the two channels, we find that CDW generally dominates near VHS filling and at higher temperatures, whereas PDW becomes favored upon detuning from the VHS and lowering the temperature.
This qualitative behavior is governed by the VHS structure and does not rely on the specific choice of coupling parameters $g$ and $v$. 
This competition between a PDW superconductivity and a CDW metal 
is reminiscent of the experimentally observed phase boundary between superconductivity and a metallic state denoted as ``undetermined metal'' \cite{HanRnGSC2025b}. 

The zero-temperature leading instabilities are shown in Fig.~\ref{fig:sus}(d).
The resulting PDW dominated regions on opposite sides of the VHS line qualitatively resemble the experimentally observed SC1 and SC2 regions.
The predicted superconducting region on the lower-left side is absent experimentally, where an insulating phase possibly associated with Wigner crystallization has instead been reported.

\textit{Discussion---}
Our results suggest a mechanism for superconductivity in R$n$G arising from purely repulsive interactions.
Near the VHS, direct PH fluctuations amplify a weak $C_3$ asymmetry and dynamically generate an effective attractive interaction. 
While CDW is favored at VHS filling, detuning the chemical potential shifts the competition toward PDW pairing, naturally producing superconducting regions on both sides of the VHS line.
Although our calculations focus on R4G, the underlying mechanism is general and is expected to apply to R5G and R6G as well, as discussed in the Supplementary Material \cite{SM}. 

While an initial $C_3$ asymmetry seed is assumed in our RG analysis, 
physically it may arise from heterostrain or spontaneous symmetry breaking of the Fermi surface. Moreover, 
asymmetric interactions can themselves induce nematic charge redistribution among the three patches \cite{halbothPomeranchuk2000}, 
potentially compete with the CDW and PDW orders.
In the Supplemental Material \cite{SM}, we present additional mean-field calculations showing that, 
when such charge redistribution is partially suppressed by an intrapatch repulsion, our qualitative conclusions remain unchanged.

Nevertheless, the relevance of noninteracting-band fermiology is suggested by recent quantum-oscillation measurements in R4G \cite{kalantreFermiology2026}, whose complex features have been proposed to originate from the noninteracting band structure when magnetic breakdown effects are taken into account \cite{zhaoMB2026}.
Despite the limitations of our approach, the qualitative correspondence between the inner VHS line and the observed SC1 and SC2 regions suggests that our VHS-based picture may provide a useful organizing principle for the superconducting phase diagram.

Due to the flat bands of R$n$G, the system is generally expected to be strongly correlated. 
Indeed, the dimensionless initial couplings used in our RG analysis are of order unity or larger. 
Our approach, however, is based on weak-coupling RG and assumes that the low-energy physics is dominated by states near the VHSs. 
It may capture the qualitative structure of the phase diagram but cannot reliably predict the transition temperatures. 
Nevertheless, the relevance of noninteracting-band fermiology is suggested by recent quantum-oscillation measurements in R4G \cite{kalantreFermiology2026}, 
whose complex features have been proposed to arise from the noninteracting band once magnetic breakdown effects were properly considered \cite{zhaoMB2026}. 
Despite these limitations, the qualitative correspondence between the inner VHS line and the observed SC1 and SC2 regions suggests that our VHS-based picture provides a useful organizing principle for the superconducting phase diagram.

\textit{Acknowledgments--} 
This work is supported by the Laboratory for Physical Sciences through its support of the Condensed Matter Theory Center at the University of Maryland. 
GPT-6 and Claude Opus 5.5 were used to assist in improving language and in verifying numerical codes and analytical derivations.
The authors are responsible for the results. 

\textit{Data availability--} 
The data supporting the findings of this article will be made publicly available upon publication.

\bibliography{refs}

\bigskip
\bigskip
\begin{center}
\textbf{\large End Matter}
\end{center}
\bigskip
\vspace{-0.2cm}

\textit{Details of the patch RG. } ---
Here we provide explicit expressions for the coefficients entering the RG equations, while a full derivation is presented in the Supplemental Material~\cite{SM}.
For a general 3-patch RG model, the 
$\alpha_{\rm ph}^e$, $\alpha_{\rm pp}$, and $\alpha_{\rm ph}^d$ 
coefficients in Eq.~\eqref{eqn:beta} are 
\begin{equation}
\begin{aligned}
    \alpha_{\rm ph}^d &=-\frac{2\pi^2}{m}\int_{\mathbf{k}}
    \partial_\ell\Pi_{\rm ph}(\xi_{c;\mathbf{k}},\xi_{c;\mathbf{k}}),\\
    \alpha_{\rm ph}^e &=-\frac{2\pi^2}{m}\int_{\mathbf{k}}
    \partial_\ell\Pi_{\rm ph}(\xi_{a;\mathbf{k}},\xi_{b;\mathbf{k}}),\\
    \alpha_{\rm pp} &=-\frac{2\pi^2}{m}\int_{\mathbf{k}}
    \partial_\ell\Pi_{\rm pp}(\xi_{a;\mathbf{k}},\xi_{b;-\mathbf{k}})~.
\end{aligned}
\end{equation}
Here, the polarization bubbles are defined as
\begin{equation}
\begin{aligned}
    \Pi_{\rm ph}(\xi_1,\xi_2) &=\frac{n_F(\xi_1)-n_F(\xi_2)}{\xi_1-\xi_2}~,\\
    \Pi_{\rm pp}(\xi_1,\xi_2) &=\frac{1-n_F(\xi_1)-n_F(\xi_2)}{\xi_1+\xi_2}~,
\end{aligned}
\end{equation}
where $n_F(\xi)=\frac{1}{1+e^{\xi/T}}$ is the Fermi-Dirac distribution.

For quadratic van Hove patches, 
\begin{equation} 
    \xi_a(\bfk)= \frac{k_{a;\parallel}^2}{2\sqrt{\eta}m}-\frac{\sqrt{\eta}k_{a;\perp}^2}{2m}~, 
\end{equation}
where $m$ is the mass and $\eta$ describes the anisotropic, 
$k_{a;\parallel} = k_x\cos \phi_a + k_y \sin\phi_a,  k_{a;\perp} = k_x\sin \phi_a - k_y \cos\phi_a$, 
all the RG coefficients can be solved exactly. 

For the direct PH term, we have 
\begin{equation}
\begin{aligned}
    \alpha_\ph^d 
    =-\frac{2\pi^2}{m}\int_{-\infty}^{\infty} d\xi\, \frac{d\rho_c(\xi)}{d\xi} \frac{\xi}{T}\frac{e^{\xi/T}}{(e^{\xi/T}+1)^2} 
    =1~,
\end{aligned}
\end{equation}
where, in the derivation, we have used $\Pi_{\ph}(\xi,\xi) = \partial n_F/\partial\xi$, $\partial_\ell n_F = -\frac{\xi}{T}\frac{e^{\xi/T}}{(e^{\xi/T}+1)^2}$, 
and the DOS of a VHS to be $\rho_c(\epsilon) = \frac{m}{2\pi^2}\log(k^2_{\mathrm{cut}}/|\epsilon|)$. 
$k_{\mathrm{cut}}$ is a momentum cutoff for the VHS to avoid divergence of the DOS. 
The direct PH contribution is therefore universal and independent of the relative geometry of the VHS patches.
We note that this contribution should also arise in a frequency-shell RG, but was omitted in Ref.~\cite{gilPDW2026}, where the corresponding ZS diagram was taken to vanish.

The exchange PH and PP contributions depend on the relative geometry of patches $a$ and $b$.
To evaluate the integrals, we introduce polar coordinates $k_x=k\cos\theta$ and $k_y=k\sin\theta$, and define $R_a(\theta)=\cos^2(\theta-\phi_a)/\sqrt{\eta}
-\sqrt{\eta}\sin^2(\theta-\phi_a)$,
such that $\xi_a(\mathbf{k})=k^2R_a(\theta)/(2m)$.
The RG coefficients then reduce to angular integrals, 
\begin{equation}\label{eqn:vhs_rg_angle}
\begin{aligned}
    \alpha_\ph^e
    &=\frac14\int_0^{2\pi}d\theta
    \frac{\operatorname{sgn}R_a(\theta) -\operatorname{sgn}R_b(\theta)} {R_a(\theta)-R_b(\theta)}~,\\
    \alpha_\pp &=-\frac14 \int_0^{2\pi}d\theta
    \frac{\operatorname{sgn}R_a(\theta+\pi) +\operatorname{sgn}R_b(\theta)} {R_a(\theta+\pi)+R_b(\theta)}~,
\end{aligned}
\end{equation}
provided that the two VHSs do not have perfect nesting. 
We note that according to Eq.~\eqref{eqn:vhs_rg_angle}, $\alpha_{\ph}^e>0$ and $\alpha_{\pp}<0$, 
consistent with repulsive and attractive interpatch interactions favoring CDW and PDW instabilities, respectively.

For three identical VHSs related by $120^\circ$ rotations, the angular integrals yield
\begin{equation}\label{eq:vhs_rg_zsp_analytic}
    \alpha_\ph^e
    =\frac{2\sqrt{\eta}}{\sqrt{3}(1+\eta)}
    \log\left|
    \frac{\sqrt{3}(1+\eta)+4\sqrt{\eta}}
         {\sqrt{3}(1+\eta)-4\sqrt{\eta}}
    \right|,~
\end{equation}
and 
\begin{equation}\label{eq:vhs_rg_bcs_analytic}
    \alpha_\pp=
    -\dfrac{8\sqrt{\eta}}{\sqrt{(\eta-3)(3\eta-1)}}
    \arctan\left[
    \dfrac{\sqrt{3}(1+\eta)-4\sqrt{\eta}} {\sqrt{(\eta-3)(3\eta-1)}} \right]~.
\end{equation}
We note the coefficient difference between our results and Ref.~\cite{gilPDW2026}. 
For $0<\eta\le1$, $\alpha=\alpha_\ph^e+\alpha_\pp>0$ when $\eta\gtrsim0.157$.
Thus, except for strongly anisotropic saddle points, exchange PH fluctuations generally dominate over PP fluctuations at VHS filling.

\begin{figure}
    \centering
    \includegraphics[width=0.99\linewidth]{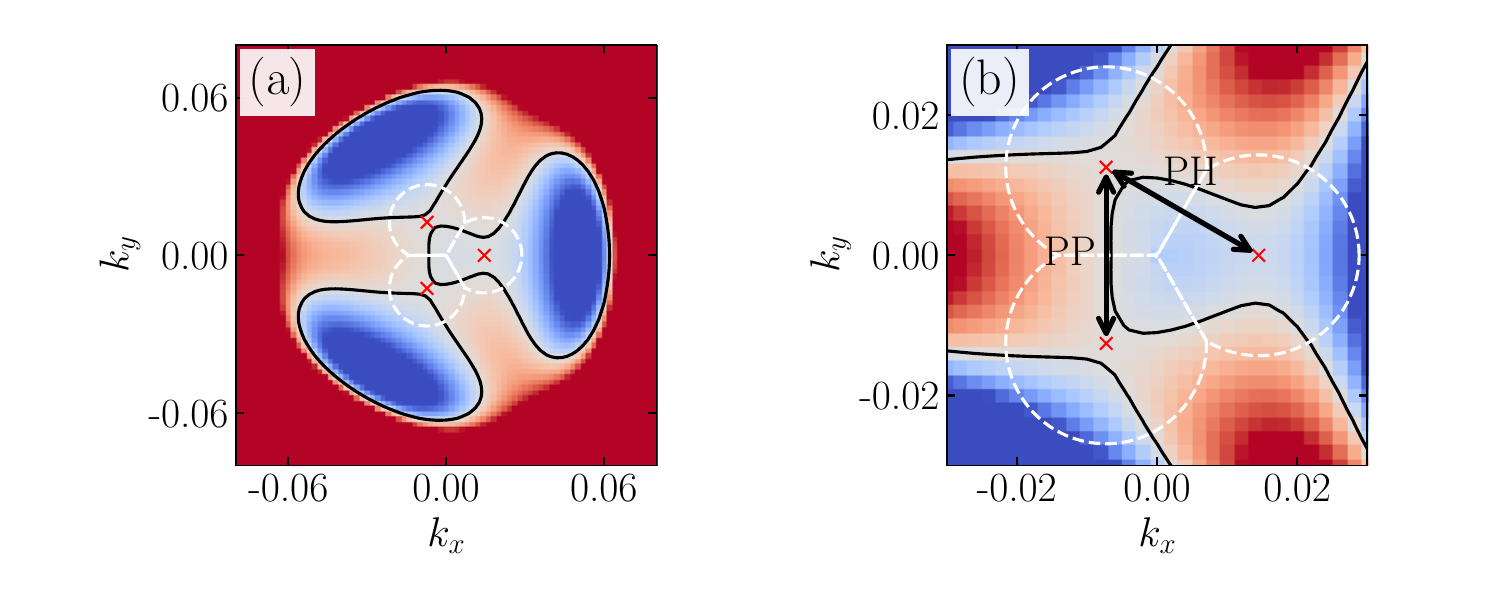}
    \caption{
    Illustration of weak $C_3$ symmetry-breaking charge redistribution of the Fermi surface.
    (a) Fermi surface geometry at $D=120$ meV and
    $n=0.36\times10^{12}\mathrm{cm}^{-2}$, with a self-energy correction $\Sigma(\bfk)=0.01\cos\theta_\bfk$ meV.
    (b) Enlarged view of (a).
    The arrows indicate PH and PP
    scattering processes between neighboring VHS patches.
    }
    \label{fig:illuC3}
\end{figure}

To identify the leading instabilities, we solve Eq.~\eqref{eqn:test_pph} near the RG divergence. 
The dominant CDW and PDW vertices scale as
$\Phi_{12}\sim\Phi_{13}\sim(\ell_c-\ell)^{-v_{\rm CDW}}$
and $\Delta_{23}\sim(\ell_c-\ell)^{-v_{\rm PDW}}$, with
\begin{equation}
    v_{\rm CDW}=\frac{(1+\alpha)\alpha_{\rm ph}^e}{\alpha^2+\alpha+1},
    \qquad
    v_{\rm PDW}=-\frac{\alpha_{\rm pp}}{\alpha^2+\alpha+1}.
\end{equation}
The corresponding susceptibilities scale as $\chi\sim(\ell_c-\ell)^{1-2v}$ and diverge for $v>1/2$.
The leading instability between CDW and PDW is therefore determined by the relative magnitudes of $(1+\alpha)\alpha_{\rm ph}^e$ and $-\alpha_{\rm pp}$.

Finally, we discuss the possibility of charge redistribution among the three VHS patches, which is generally expected when the interactions break $C_3$ symmetry.
We introduce source fields $\rho_a$ coupled to the patch densities through $\rho_a\psi_a^\dagger\psi_a$.
Their one-loop RG equations read
\begin{equation}
\begin{aligned}
    \frac{d\rho_a}{d\ell}
    &=-\frac{m}{2\pi^2}\alpha_{\rm ph}^d
    \sum_{b\neq a}g_{ab}\rho_b.
\end{aligned}
\end{equation}
The three eigenmodes of the charge vertices scale as
$\rho_i\sim(\ell_c-\ell)^{-v_i}$, with exponents
\begin{equation}
    v_\pm = \frac{1\pm\sqrt{1+8(1+\alpha)^2}} {2(\alpha^2+\alpha+1)},\quad
    v_0 =-\frac{1}{\alpha^2+\alpha+1}~.
\end{equation}
The leading exponent $v_+$ can exceed both $v_{\rm CDW}$ and $v_{\rm PDW}$.

\textit{$C_3$ charge redistribution. } ---
The test-vertex analysis suggests that $C_3$ symmetry-breaking charge redistribution can emerge as the leading instability. 
Despite the strong enhancement of the $C_3$-breaking charge instability, its effect on the CDW and PDW orders may be limited for several reasons.

First, our simplified patch model neglects intrapatch interactions due to the Pauli principle.
For finite-sized patches, however, residual momentum-dependent intrapatch interactions can remain and may suppress charge redistribution.
Moreover, the patches are connected to the rest of the Fermi surface, which imposes additional constraints on charge redistribution.

Second, a moderate $C_3$-breaking charge redistribution does not necessarily destroy the scattering geometry discussed in Fig.~\ref{fig:illu}(c). 
As illustrated in Fig.~\ref{fig:illuC3}, when patch 1 is electron-doped relative to its VHS while patches 2 and 3 are hole-doped, 
the resulting geometry can still favor PH scattering between patches 1 and 2 and PP scattering between patches 2 and 3.
The competition among these orders is further examined within mean-field analysis in the Supplemental Material \cite{SM}.

\newpage \clearpage 
	
\onecolumngrid
	
\begin{center}
	{\large
        Emergent Attraction and Competing Orders near the Van Hove Singularity in Rhombohedral $n$-layer Graphene
		\vspace{4pt}
		\\
		SUPPLEMENTAL MATERIAL
	}
\end{center}
	
\setcounter{figure}{0}
\renewcommand{\thefigure}{S\arabic{figure}}
\setcounter{equation}{0}
\renewcommand{\theequation}{S\arabic{equation}}


\section{Realistic band parameters. }

We describe rhombohedral $n$-layer graphene using the continuum Hamiltonian
\begin{equation}
    H_0=\sum_{\eta=\pm1}\sum_{\mathbf{k}}
    \left[ \sum_{l=1}^{n} \Psi_{\eta l\mathbf{k}}^\dagger
        h_l(\mathbf{k}) \Psi_{\eta l\mathbf{k}}
    + \sum_{l=1}^{n-1} \Psi_{\eta l\mathbf{k}}^\dagger T_1(\mathbf{k}) \Psi_{\eta,l+1,\mathbf{k}} +\mathrm{h.c.}
    + \sum_{l=1}^{n-2} \Psi_{\eta l\mathbf{k}}^\dagger T_2 \Psi_{\eta,l+2,\mathbf{k}}
    +\mathrm{h.c.}
    \right],
\end{equation}
where $\eta=\pm1$ labels the $K$ and $K'$ valleys,
$\mathbf{k}$ is measured relative to the corresponding valley center, 
and $\Psi_{\eta l\mathbf{k}}=(c_{A_l},c_{B_l})^T$
is the two-component spinor on layer $l$.
The intralayer and interlayer hopping matrices are
\begin{equation}
    h_l(\mathbf{k}) = \begin{pmatrix}
    V_l & v_0\Pi_\eta^\dagger\\
    v_0\Pi_\eta & V_l \end{pmatrix} +U_l, 
    \qquad
    T_1(\mathbf{k}) = \begin{pmatrix}
        v_4\Pi_\eta^\dagger & v_3\Pi_\eta\\
        \gamma_1 & v_4\Pi_\eta^\dagger \end{pmatrix},
    \qquad
    T_2= \begin{pmatrix}
        0 & \gamma_2/2\\
        0 & 0
    \end{pmatrix},
\end{equation}
where $\Pi_\eta=\eta k_x+ik_y$ and
$v_j=\sqrt{3}\gamma_j a/2$, with
$a=2.46\,\text{\AA}$ the graphene lattice constant.

The displacement field introduces a layer-dependent
potential
\begin{equation}
    V_l=\left(l-\frac{n+1}{2}\right)\frac{D}{(n-1)},
\end{equation}
where $D$ is the potential difference between
the top and bottom layers.
The additional onsite potentials are 
\begin{equation}
    U_l= \begin{pmatrix}
    \Delta_l+\delta(1-\delta_{l,1}) & 0\\
    0 & \Delta_l+\delta(1-\delta_{l,n})
    \end{pmatrix},
\end{equation}
with $\Delta_l=\Delta_2$ for $l=1,n$ and
$\Delta_l=-\Delta_2$ otherwise.
This parametrization reproduces the onsite potentials
of rhombohedral tetralayer graphene.

We focus on the $K$ valley and adopt the hopping
parameters of Refs.~\cite{kalantreFermiology2026,auerbachIsospin2025}:
$\gamma_0=3100$ meV, $\gamma_1=380$ meV,
$\gamma_2=-15$ meV, $\gamma_3=-290$ meV,
and $\gamma_4=-141$ meV, together with
$\delta=10.5$ meV and $\Delta_2=2$ meV.

\section{Extension to general R$n$G}

\begin{figure}[t]
    \centering
    \includegraphics[width=0.9\linewidth]{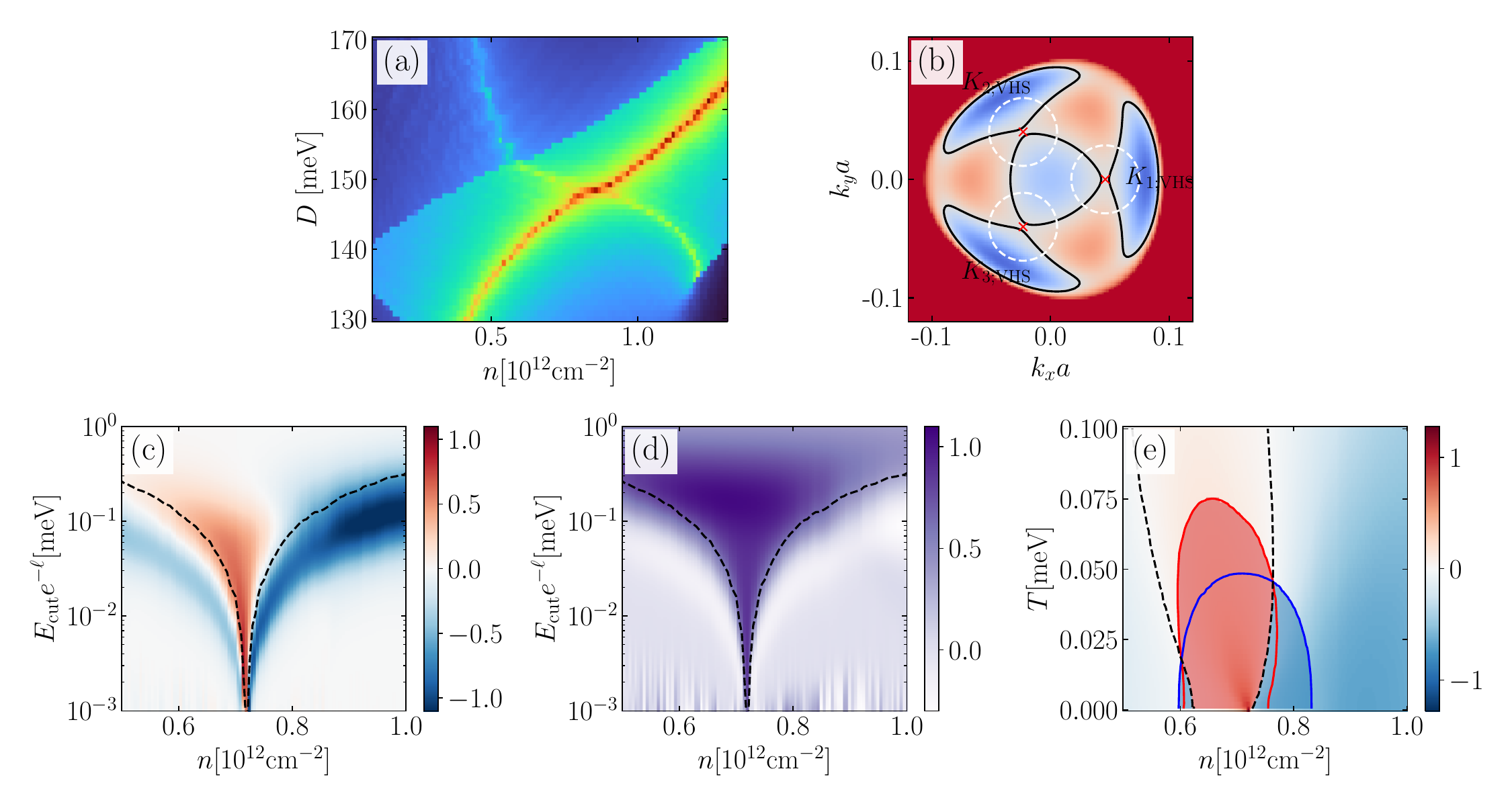}
    \caption{Patch RG and susceptibilities of R6G.
    (a) DOS as a function of carrier density $n$ and displacement field $D$ between the top and bottom layers.
    The VHS peaks exhibit an X-shaped pattern similar to that in Fig.~\ref{fig:illu}.
    The branch extending from the lower left to the upper right corresponds to the inner VHSs connecting the inner and outer Fermi pockets, 
    while the other branch corresponds to the outer VHSs associated with the evolution of an annular Fermi surface into three outer pockets.
    (b) Representative Fermi surface at $D=145$ meV and $n=0.69\times10^{12}\mathrm{cm}^{-2}$.
    The three inner VHSs are marked by red crosses, and the corresponding patches by white dashed circles with radius $k_{\mathrm{cut}}a=0.03$. 
    (c) RG coefficient $\alpha=\alpha_\ph^e+\alpha_\pp$ as a function of $n$ and energy scale $E=E_{\rm cut}e^{-\ell}$ at $D=145$ meV.
    The dashed line indicates the detuning scale $|\mu-E_{\VHS}|$.
    (d) Direct PH coefficient $\alpha_\ph^d$ for the same parameters as in (c).
    (e) Difference between the CDW and PDW susceptibilities weighted by their respective interactions, $-g\int_\bfk\Pi_\ph(\xi_{1;\bfk},\xi_{2;\bfk})-v\int_\bfk\Pi_\pp(\xi_{2;\bfk},\xi_{3;-\bfk})$, 
    with $g=550$ meV$\cdot$nm$^2$ and $v=550$ meV$\cdot$nm$^2$. 
    The red and blue curves indicate the CDW and PDW mean-field instability boundaries, respectively, where the corresponding weighted susceptibilities reach unity.
    The dashed line marks their equality.
    The regions dominated by CDW and PDW instabilities are shaded in red and blue, respectively.
    }
    \label{fig:R6G}
\end{figure}

The physical picture discussed in the main text also applies to R5G and R6G.
As an additional example, Fig.~\ref{fig:R6G} presents the corresponding analysis for R6G. 
Similar to R4G, the DOS exhibits an X-shaped pattern of VHS peaks, with three inner VHSs connecting the inner and outer Fermi pockets.

Focusing on the inner VHS, the patch RG coefficients show the same crossover from PH-dominated fluctuations near VHS filling to PP-dominated fluctuations at low energy scales upon detuning.
The susceptibility analysis also shows leading CDW instability near the VHS and leading PDW instabilities on either side of VHS. 
These results suggests that the proposed mechanism is not specific to R4G but extends generally to rhombohedral $n$-layer graphene with $n=4,5,6$. 

\section{Patch mean-field analysis}

In this section, we provide a mean-field analysis of the patch model. 
We consider the same spinless fermions in three patches around the $C_3$-related saddle points
$\mathbf K_a$ ($a=1,2,3$), with a circular momentum cutoff $|\mathbf k|<k_c$ around each saddle point. 
When neighboring disks overlap, each state is assigned to the patch whose saddle point is closest. 
\begin{equation}\label{eqn:patch_int_app}
    H_{\rm int} = \int d^2 r\left( g\norm{\hat n_1\hat n_2}+g\norm{\hat n_1\hat n_3}
    -v\norm{\hat n_2\hat n_3}\right)
    +\frac{\delta g}{2\mathcal A}\sum_{a=1}^{3}\norm{\hat N_a^{2}}~,
\end{equation}
where $g,v>0$, $\hat n_a(\mathbf r)=\psi_a^\dagger(\mathbf r)\psi_a(\mathbf r)$ is the local density of patch $a$,
$\hat N_a=\int d^2r\hat n_a(\mathbf r)$ is the total number of electrons in patch $a$, and $\mathcal A$ is the system area. 
Here, normal ordering is defined with respect to the noninteracting Fermi sea at the reference chemical potential. 
The intrapatch forward-scattering interaction $\delta g>0$ accounts for residual Coulomb repulsion within finite-sized patches, 
which arises from the momentum dependence of a finite-range interaction $u(\mathbf q)$,
$\delta g\simeq u(0)-\langle u(\mathbf k-\mathbf k')\rangle_{\mathbf k,\mathbf k'\in a}$.
In a realistic system, the patches are connected through the underlying band structure, and $C_3$ symmetry breaking is further constrained by the electronic states outside the patches.

We consider the CDW and PDW order parameters
\begin{equation}
\Phi =g\int_{\bf k}
\langle c^\dagger_{2;\bf k}c_{1;\bf k}\rangle,\qquad
\Delta =-v\int_{\bf k}
\langle c_{3;-\bf k}c_{2;\bf k}\rangle~.
\end{equation}
We also allow for $C_3$ symmetry breaking through self-consistent Hartree shifts,
\begin{equation}
\Sigma_1=\delta g\delta n_1+g\delta n_2+g\delta n_3,\qquad
\Sigma_2=g\delta n_1+\delta g\delta n_2-v\delta n_3,\qquad
\Sigma_3=g\delta n_1-v\delta n_2+\delta g\delta n_3~,
\end{equation}
where $\delta n_a=n_a-n_a^{(0)}$ denotes the density deviation from the noninteracting reference state.
The intrapatch forward-scattering term contributes to the Hartree shifts, while its exchange contribution is suppressed by $1/\mathcal A$.

The mean-field Hamiltonian is
\begin{equation} \label{eq:Kmf}
    H_{\rm MF} =\int_\bfk
    \Bigl[\Psi_{\bfk}^\dagger A(\bfk)\Psi_{\bfk}+\xi_3(-\bfk)\Bigr]~,
\end{equation}
where $\xi_a(\bfk)=\epsilon_a(\bfk)+\Sigma_a-\mu$. 
And 
\begin{equation} \label{eq:bdg}
    \Psi_{\bfk}=\begin{pmatrix}c_{1,\bfk}\\ c_{2,\bfk}\\ c^\dagger_{3,-\bfk}\end{pmatrix},
    \qquad
    A(\bfk)=\begin{pmatrix}
    \xi_1(\bfk)&-\Phi&0\\
    -\Phi&\xi_2(\bfk)&\Delta\\
    0&\Delta&-\xi_3(-\bfk)
    \end{pmatrix}~. 
\end{equation}

\begin{figure}[t]
    \centering
    \includegraphics[width=0.85\linewidth]{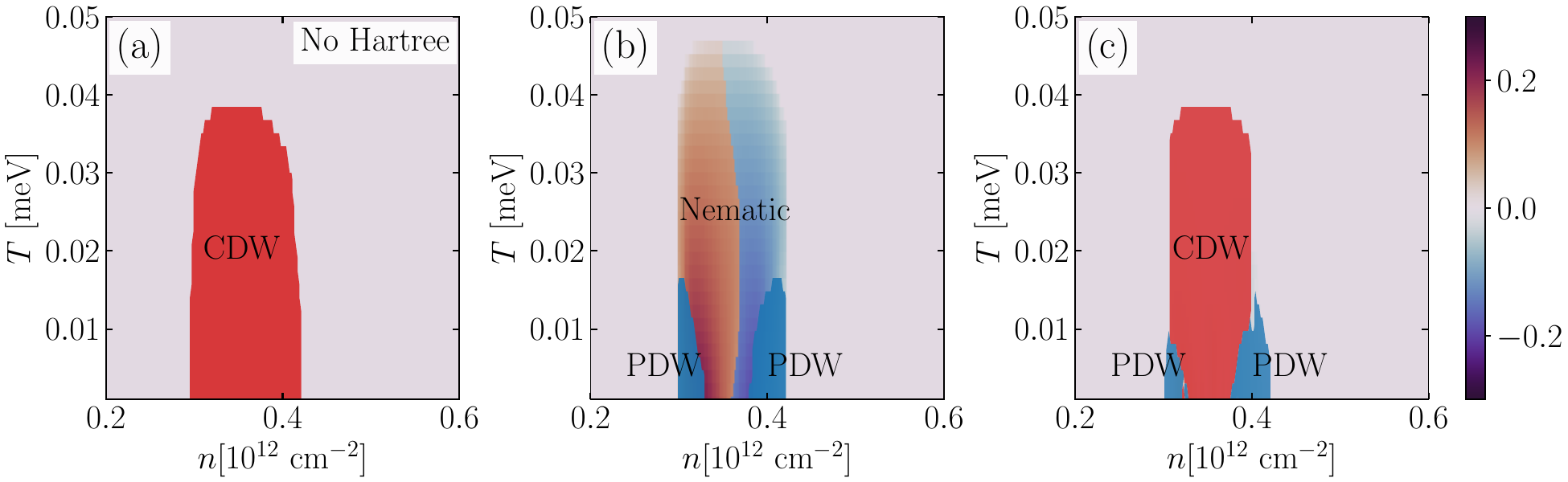}
    \caption{
    Mean-field phase diagrams of the patch model defined in Eq.~\eqref{eqn:patch_int_app} for R4G, with $g=608$ meV$\cdot$nm$^2$, $v=305$ meV$\cdot$nm$^2$, and $k_{\mathrm{cut}}a=0.015$.
    (a) Mean-field phase diagram without Hartree corrections. 
    (b), (c) Mean-field phase diagrams including self-consistent Hartree shifts and $C_3$-breaking charge redistribution.
    The background color indicates the Hartree shift $\Sigma_1$.
    Finite intrapatch repulsions $\delta g=g/2$ in (b) and $\delta g=g$ in (c) are introduced to suppress charge redistribution.
    }
    \label{fig:MF}
\end{figure}

We obtain self-consistent solutions from different initial seeds and identify the energetically favored state by comparing their grand potentials at fixed chemical potential.
The grand potential is given by
$\Omega=\langle H-H_{\MF}\rangle_{\MF}-T\log Z_{\MF}$, yielding
\begin{equation} \label{eq:F}
    \Omega=\int_\bfk\Bigl\{\xi_3(-\bfk)-T\sum_{\lambda=1}^{3}
    \ln\bigl[1+e^{-E_\lambda(\bfk)/T}\bigr]\Bigr\}
    +\frac{\Phi^2}{g}+\frac{\Delta^2}{v} 
    -\sum_a\left(\frac12\delta n_a+n_0\right)\Sigma_a~,
\end{equation}
where $E_\lambda(\bfk)$ are the eigenvalues of $A(\bfk)$, $\delta\bm n=(\delta n_1,\delta n_2,\delta n_3)^T$, and $n_0$ is the noninteracting reference density per patch.

The mean-field phase diagrams are shown in Fig.~\ref{fig:MF}.
We use a weaker interaction strength than in Fig.~\ref{fig:sus} to avoid fully gapping out the patches.
We assume that electronic states outside the patches act as an effective charge reservoir, allowing the chemical potential to remain fixed during the mean-field iterations.
Since our analysis is intended to be qualitative, this assumption is not considered essential to our conclusions.
The chemical potential is determined from the density dependence $\mu(n)$ of the full noninteracting band structure.

When the Hartree shifts $\Sigma_a$ are neglected, the resulting phase diagram, shown in Fig.~\ref{fig:MF}(a) 
is dominated by the CDW order. 
Although the susceptibility analysis favors PDW at low temperatures away from the VHS, the CDW state has a lower free energy and thus preempts the PDW phase.
When Hartree shifts $\Sigma_a$ are considered, a pronounced $C_3$-breaking charge imbalance can indeed develop.
For a relatively weak intrapatch repulsion $\delta g=g/2$, the charge redistribution is strong, and 
CDW order is suppressed near VHS filling.
On the other hand, the PDW order away from the VHS filling survives. 
For a stronger intrapatch repulsion $\delta g=g$, the charge imbalance remains finite but no longer suppresses CDW order.

In realistic systems, both $C_3$-breaking charge redistribution and intrapatch repulsion are expected to be physically relevant. 
Our results show that, when both effects are taken into account, the qualitative picture proposed in the main text can still be realized, with PDW superconductivity emerging on both sides of the VHS.
However, the nature of the high-temperature metallic state becomes more subtle, as it may be dominated by either CDW order or $C_3$-breaking nematic charge redistribution.
This provides two possible interpretations of the experimentally observed ``undetermined metal'' \cite{HanRnGSC2025b}.
Nevertheless, PDW order remains robust upon detuning from the VHS, leaving our main conclusion of superconductivity on both sides of the VHS essentially unchanged.

\section{Temperature-flow renormalization group equations of the VHSs}
\label{app:tflow}

In this appendix, we briefly review the temperature-flow renormalization group (RG) approach to the patch VHS model, following Honerkamp and Salmhofer~\cite{honerkampTfRG2001}.
A key advantage of temperature-flow RG is that it treats low-energy direct particle-hole screening and other instabilities on an equal footing.
This is particularly important near a VHS, where particle-hole fluctuations are logarithmically enhanced and can significantly renormalize the effective interactions.
In contrast, such screening effects are missed by the conventional momentum-shell Wilsonian RG~\cite{shankarRG1994}, but can be captured by a frequency-shell RG scheme.

\subsection{General temperature RG}

Consider a general action of a spinless Fermion
\begin{equation} \label{eqn:tflow_action}
    S = T\int_k \bar\psi_k(-i\omega_n+\xi_{\bfk})\psi_k
    +\frac{T^3}{4}\int_{k_1,k_2,k_3}u(\bfk_4,\bfk_3,\bfk_2,\bfk_1)
    \bar\psi_{k_4}\bar\psi_{k_3}\psi_{k_2}\psi_{k_1} 
    + \int_k (\bar{J}_k\psi_k + \bar{\psi}_kJ_k)~,
\end{equation}
where $\omega_n = (2n+1)\pi T$, $k=(n,\bfk)$, $\psi_k = \psi_{n;\bfk}$ and $\int_k\equiv \sum_n \int \frac{d^2k}{(2\pi)^2}$. 
Due to energy-momentum conservation, $k_4=k_1+k_2-k_3$. 

To get RG equations, replace the quadratic kernel with $T(-i\omega_n+\xi_{\mathbf{k}})\to Q_\Lambda(k)$, which depends on a continuous parameter $\Lambda$. 
For every $\Lambda$, define
\begin{equation}\label{eqn:tflow_ZW}
    Z_\Lambda[\bar J,J]=\int\mathcal D\bar\psi\,\mathcal D\psi\,
    e^{-S[Q_\Lambda]},\qquad W_\Lambda=-\ln Z_\Lambda,
\end{equation}
from which we can get the mean field parameters $ \phi_k=\langle\psi_k\rangle_J=\frac{\delta W}{\delta\bar J_k}$, 
$\bar\phi_k=\langle\bar\psi_k\rangle_J=-\frac{\delta W}{\delta J_k}$. 
At fixed sources, differentiating $W_\Lambda$ gives
\begin{equation}\label{eqn:tflow_dW}
    \partial_\Lambda W|_J = \int_k\partial_\Lambda Q_\Lambda(k)\langle\bar\psi_k\psi_{k}\rangle_J
    =\int_k\partial_\Lambda Q_\Lambda(k) \bar\phi_k\phi_{k}
    -\mathrm{Tr}\partial_\Lambda Q_\Lambda G_\Lambda[J].
\end{equation}
where we introduced the connected Green's function as 
$G_\Lambda[J]=\langle\psi\bar\psi\rangle_J-\phi\bar\phi$.

We now do the Legendre transform to get the effective action
\begin{equation}
    \Gamma_\Lambda[\bar\phi,\phi]
    =W_\Lambda[\bar J,J]-\int_k(\bar J_k\phi_k+\bar\phi_kJ_k).
\end{equation}
where $J_k$ is an implicily function of $\phi_k$ through $\phi_k = \delta W/\delta\bar J_k$. 
It is straightforward to show 
\begin{equation} \label{eqn:tflow_dGamma}
    \partial_\Lambda\Gamma_\Lambda|_{\phi}=\partial_\Lambda W_\Lambda|_{J}+ 
    \int_k\partial_\Lambda\bar J_k\frac{\delta W}{\delta\bar J_k} +\int_k\partial_\Lambda J_k\frac{\delta W}{\delta J_k}
    -\int_k(\partial_\Lambda\bar J_k \phi_k+\bar\phi_k\partial_\Lambda J_k)
    =\partial_\Lambda W_\Lambda|_J~.
\end{equation}
The last four terms cancel pairwise as  $\delta W/\delta\bar J=\phi$ and $\delta W/\delta J=-\bar\phi$. 
Inserting Eq.~\eqref{eqn:tflow_dW} into Eq.~\eqref{eqn:tflow_dGamma} gives
\begin{equation}\label{eqn:Gammaflow}
    \partial_\Lambda\Gamma_\Lambda
    =\int_k\partial_\Lambda Q_\Lambda(k)\bar\phi_k\phi_k
    -\operatorname{Tr}\left[\partial_\Lambda Q_\Lambda\,G_\Lambda[\bar\phi,\phi]\right],
\end{equation}
where \(G_\Lambda[\bar\phi,\phi]\) is the field-dependent connected Green's function, obtained from the inverse Legendre relation between the connected two-point functions and the full Hessian of \(\Gamma_\Lambda\).

To derive the RG equations, we expand the effective action as
\begin{equation}\label{eqn:tflow_vertexexp}
    \Gamma_\Lambda[\bar\phi,\phi]
    =\Gamma_\Lambda^{(0)}
    +\int_k (Q_\Lambda(k)+\Sigma_\Lambda(k))\bar\phi_k\phi_k
    +\frac{T^3}{4}\int_{k_1,k_2,k_3}
    u_\Lambda(k_4, k_3, k_2, k_1)
    \bar\phi_{k_4}\bar\phi_{k_3}\phi_{k_2}\phi_{k_1}
    +O(\phi^6).
\end{equation}

Comparing the left and right-hand sides of Eq.~\eqref{eqn:Gammaflow}, we get the flow equation of the self-energy: 
\begin{equation}\label{eqn:tflow_sigma}
    \partial_\Lambda\Sigma_\Lambda(k) =-\int_{k'} S_\Lambda(k') u_\Lambda(k',k,k,k')~,
\end{equation}
and the flow equation of the vertex as 
\begin{equation}\label{eqn:tflow_du}
    \partial_\Lambda u_\Lambda(k_4,k_3,k_2,k_1) =\delta u_{\mathrm{ph}}^d + \delta u_{\mathrm{ph}}^e + \delta u_{\mathrm{pp}}~,
\end{equation}
where it is decomposed into 
the direct particle-hole fluctuation $\delta u_{\mathrm{ph}}^d$
the exchange particle-hole fluctuation $\delta u_{\mathrm{ph}}^e$
and the particle-particle fluctuation $\delta u_{\mathrm{pp}}$: 
\begin{equation}
\begin{aligned}
    \delta u_{\mathrm{ph}}^d &=+\int_k L_\Lambda(k+k_4,k+k_1)
    u_\Lambda(k_4,k+k_1,k+k_4,k_1) u_\Lambda(k+k_4,k_3,k_2,k+k_1),\\
    \delta u_{\mathrm{ph}}^e &=-\int_k L_\Lambda(k+k_2,k+k_4)
    u_\Lambda(k+k_4,k_3,k+k_2,k_1) u_\Lambda(k_4,k+k_2,k_2,k+k_4),\\
    \delta u_{\mathrm{pp}} &=-\frac12\int_k L_\Lambda(k+k_1,k_2-k)
    u_\Lambda(k_4,k_3,k_2-k,k+k_1) u_\Lambda(k+k_1,k_2-k,k_2,k_1)~,
\end{aligned}
\end{equation}

Here, the single-line propagator is defined as 
\begin{equation} \label{eqn:tflow_S}
    S_\Lambda=-G_\Lambda\partial_\Lambda Q_\Lambda G_\Lambda,  
\end{equation}
and the double-line kernel
\begin{equation} \label{eqn:tflow_L}
    L_\Lambda(k,k') =S_\Lambda(k)G_\Lambda(k')+G_\Lambda(k)S_\Lambda(k').
\end{equation}

\subsection{Temperature as the flow parameter}
\label{app:tflow:rescaling}

The framework above requires that only the quadratic kernel depends on the flow parameter. 
Following Ref.~\cite{honerkampTfRG2001} we introduce $\eta_k=T^{3/4}\psi_k, 
    \bar\eta_k=T^{3/4}\bar\psi_k,
    \tilde J_k=T^{-3/4}J_k,
    \tilde{\bar J}_k=T^{-3/4}\bar J_k$, 
such that only the quadratic term depends on the temperature as 
$Q_{T}(k)=T^{-1/2}\bigl(-i(2n+1)\pi T+\xi_{\mathbf{k}}\bigr)$, 
and all results above apply with mapping $\Lambda\to T$.
The single and double line propagators  Eqs.~\eqref{eqn:tflow_S} and \eqref{eqn:tflow_L} are
\begin{equation}
    S_T(\mathbf{k},n)
    =\frac{T^{-1/2}}{2}\,\frac{i\omega_n+\xi_{\mathbf{k}}}{(-i\omega_n+\xi_{\mathbf{k}})^2},\qquad 
    L_T(k,k') 
    =\frac{\mathrm{d}}{\mathrm{d}T}\left[\frac{T}{\bigl(-i\omega_n+\xi_{\mathbf{k}}\bigr)\bigl(-i\omega_{n'}+\xi_{\mathbf{k}'}\bigr)}\right]~.
\end{equation}

Take $\ell = \log (E_{\mathrm{cut}}/T)$ as the scaling time, and omit the flow of self-energy as well as the frequency dependence of $u$, 
the RG equations read as 
\begin{equation}
    \partial_\ell u(\bfk_4,\bfk_3,\bfk_2,\bfk_1)= \delta  u_{\mathrm{ph}}^d+\delta u_{\mathrm{ph}}^e+\delta u_{\mathrm{pp}},
\end{equation}
where 
\begin{equation}
\begin{aligned}
    \delta u_{\mathrm{ph}}^d &=\int_{\bfk}\partial_\ell \Pi_{\mathrm{ph}}(\xi_{\bfk+\bfk_4},\xi_{\bfk+\bfk_1}) u(\bfk_4,\bfk+\bfk_1,\bfk+\bfk_4,\bfk_1)u(\bfk+\bfk_4,\bfk_3,\bfk_2,\bfk+\bfk_1),\\
    \delta u_{\mathrm{ph}}^e&=-\int_{\bfk}\partial_\ell \Pi_{\mathrm{ph}}(\xi_{\bfk+\bfk_2},\xi_{\bfk+\bfk_4})
    u(\bfk_4,\bfk+\bfk_2,\bfk_2,\bfk+\bfk_4)u(\bfk+\bfk_4,\bfk_3,\bfk+\bfk_2,\bfk_1),\\
    \delta u_{\mathrm{pp}} &=-\frac12\int_{\bfk}\partial_\ell \Pi_{\mathrm{pp}}(\xi_{\bfk+\bfk_1},\xi_{\bfk_2-\bfk})
    u(\bfk_4,\bfk_3;\bfk_2-\bfk, \bfk+\bfk_1)u(\bfk+\bfk_1,\bfk_2-\bfk,\bfk_2,\bfk_1)~,
\end{aligned}
\end{equation}
and the polarizations are defined as 
\begin{align}
    \Pi_{\mathrm{ph}}(\xi_1,\xi_2) &=T\sum_n\frac{1}{i\omega_n-\xi_1}\frac{1}{i\omega_n-\xi_2} =\frac{n_F(\xi_1)-n_F(\xi_2)}{\xi_1-\xi_2},\\
    \Pi_{\mathrm{pp}}(\xi_1,\xi_2) &=T\sum_n\frac{1}{i\omega_n-\xi_1}\frac{1}{-i\omega_n-\xi_2} =\frac{1-n_F(\xi_1)-n_F(\xi_2)}{\xi_1+\xi_2}.
\end{align}
$n_F(\xi) = 1/(e^{\xi/T}+1)$ is the Fermi-Dirac distribution.

\subsection{Test-vertex flow}

To investigate the competing particle-hole and particle-particle instabilities, we introduce infinitesimal external sources coupled to the corresponding bilinears,
\begin{equation}\label{eq:source}
    \delta S=\int_k\left[
    \Phi_{\mathbf k,\mathbf Q}\bar\psi_{k+Q}\psi_k
    +\frac12\Delta_{\mathbf k,\mathbf Q}
    \bar\psi_k\bar\psi_{Q-k}
    +\mathrm{c.c.}\right],
\end{equation}
where $Q=(0,\mathbf Q)$ denotes the ordering momentum.
The pairing source satisfies
$\Delta_{\mathbf k,\mathbf Q}=-\Delta_{\mathbf Q-\mathbf k,\mathbf Q}$
due to fermionic antisymmetry.

At the one-loop level, the renormalized test vertices obey
\begin{equation}\label{eqn:ph_test}
    \partial_\ell\Phi_{\mathbf k,\mathbf Q}
    =\int_{\mathbf k'}
    u(\bfk+\bfQ,\bfk',\bfk'+\bfQ,\bfk)
    \partial_\ell\Pi_{\mathrm{ph}}
    (\xi_{\mathbf k'},\xi_{\mathbf k'+\mathbf Q})
    \Phi_{\mathbf k',\mathbf Q},
\end{equation}
and
\begin{equation}\label{eqn:pp_test}
    \partial_\ell\Delta_{\mathbf k,\mathbf Q}
    =-\frac12\int_{\mathbf k'}
    u(\bfk,\bfQ-\bfk,\bfQ-\bfk',\bfk')
    \partial_\ell\Pi_{\mathrm{pp}}
    (\xi_{\mathbf k'},\xi_{\mathbf Q-\mathbf k'})
    \Delta_{\mathbf k',\mathbf Q}.
\end{equation}

\section{One-loop RG for quadratic Van Hove patches}

We now derive the one-loop renormalization group for a set of quadratic van Hove patches with momenta $\mathbf{K}_{a;\VHS}$ 
\begin{equation} \label{eqn:VHS_e}
    \xi_a(\bfk)= \frac{k_{a;\parallel}^2}{2\sqrt{\eta}m}-\frac{\sqrt{\eta}k_{a;\perp}^2}{2m}~, 
\end{equation}
where $m$ is the mass and $\eta$ describes the anisotropic. 
The relative momentum $\bfk$ is measured from the corresponding VHS center $\mathbf{K}_{a;\VHS}$, and 
$k_{a;\parallel} = k_x\cos \phi_a + k_y \sin\phi_a, 
k_{a;\perp} = k_x\sin \phi_a - k_y \cos\phi_a$.

We consider the interaction
\begin{equation} \label{eqn:Hint}
    H_{\mathrm{int}}=\sum_{a<b} g_{ab} \int \mathrm{d}^2x\, \bar\psi_a\bar\psi_b\psi_b\psi_a~,
\end{equation}
which can be extracted by projecting the real interaction near the VHSs, $g_{ab} = u(\bfK_{a;\VHS}, \bfK_{b;\VHS}, \bfK_{b;\VHS}, \bfK_{a;\VHS})$. 
All the interactions within a VHS patch are assumed to be constant.
The intra-patch interaction $g_{aa}=0$ vanishes due to the Pauli principle. 
Following Ref.~\cite{honerkampTfRG2001}, we employ the temperature-flow RG with $\ell=\log(E_{\mathrm{cut}}/T)$. The one-loop RG equations for the interpatch couplings are
\begin{equation}\label{eqn:beta_app}
\begin{aligned}
    \partial_\ell{g_{ab}} =& \delta g_{ab}|_\ph^d + \delta g_{ab}|_\ph^e + \delta g_{ab}|_\pp\\
    =&\sum_{c\neq a,b}\int_\bfk \partial_\ell \Pi_{\mathrm{ph}}(\xi_{c;\bfk},\xi_{c;\bfk})g_{ac}g_{bc}
    -\int_\bfk\left[
    \partial_\ell \Pi_{\mathrm{ph}}(\xi_{a;\bfk},\xi_{b;\bfk})
    +\partial_\ell \Pi_{\mathrm{pp}}(\xi_{a;\bfk},\xi_{b;-\bfk})
    \right]g_{ab}^2~. 
\end{aligned}
\end{equation}
Defining 
\begin{equation}
\begin{aligned}
    \alpha_\ph^d=-\frac{2\pi^2}{m}\int_\bfk\partial_\ell\Pi_{\mathrm{ph}}(\xi_{c;\bfk},\xi_{c;\bfk}),\quad
    \alpha_{\ph}^e=-\frac{2\pi^2}{m}\int_\bfk\partial_\ell\Pi_{\mathrm{ph}}(\xi_{a;\bfk},\xi_{b;\bfk}),\quad
    \alpha_\pp=-\frac{2\pi^2}{m}\int_\bfk\partial_\ell\Pi_{\mathrm{pp}}(\xi_{a;\bfk},\xi_{b;-\bfk})
\end{aligned}
\end{equation}
The flow equation Eq.~\eqref{eqn:beta_app} reduces to Eq.~\eqref{eqn:beta} in the main text. 

Here, the direct particle-hole contribution involves density fluctuations mediated through the same third patch. 
This contribution therefore requires at least three VHS patches and vanishes in a two-patch model.
For repulsive interactions $g_{ac},g_{bc}>0$, it gives a negative correction to $g_{ab}$, analogous to the screening of the repulsive interaction by density fluctuations in the third patch.
The exchange particle-hole and particle-particle contributions, in contrast, depend on the relative geometry of patches $a$ and $b$.

To identify the leading instabilities, we introduce test vertices in the direct particle-hole $\rho_a\psi^\dagger_a\psi_a$, exchange particle-hole $\Phi_{ab}\psi^\dagger_a\psi_b$, and particle-particle $\Delta_{ab}\psi_a\psi_b$ channels, 
corresponding to charge redistribution, CDW, and PDW orders, respectively. 
Their RG flow equations are given by Eqs.~\eqref{eqn:ph_test} and \eqref{eqn:pp_test} as 
\begin{align}
    \frac{d\rho_a}{d\ell} &=-\frac{m}{2\pi^2}\alpha_{\rm ph}^d \sum_{b\neq a}g_{ab}\rho_b~,\\
    \frac{d\log\Phi_{ab}}{d\ell}
    &=\frac{m}{2\pi^2}\alpha_{\ph}^eg_{ab}(\ell)~, \\
    \frac{d\log\Delta_{ab}}{d\ell}
    &=\frac{m}{2\pi^2}\alpha_{\pp}g_{ab}(\ell)~.\\ 
\end{align}

\section{Flow fixed points of the RG equation. }

In this section, we analyze the flow of the beta functions in Eq.~\eqref{eqn:beta} at the VHS energy. 
For later convenience, we take the dimensionless parameter $\lambda_1 = \frac{m}{2\pi^2}g_{23}$, 
$\lambda_2 = \frac{m}{2\pi^2}g_{13}$ and $\lambda_3 = \frac{m}{2\pi^2}g_{12}$. 
The flow equation Eq.~\eqref{eqn:beta} then becomes 
\begin{equation} \label{eqn:triflow}
    \frac{d \lambda_1}{d\ell}=\alpha\lambda_1^2 - \lambda_2\lambda_3, \qquad
    \frac{d\lambda_2}{d\ell}=\alpha\lambda_2^2 - \lambda_3\lambda_1, \qquad
    \frac{d\lambda_3}{d\ell}=\alpha\lambda_3^2 - \lambda_1\lambda_2~.
\end{equation}

\subsection{Exact solutions for $\alpha=0$}

We first discuss the analytically solvable case $\alpha=0$.
Without loss of generality, we consider the initial values  $|\lambda_1(0)|\geq|\lambda_2(0)|\geq|\lambda_3(0)|\geq 0$,
Since the flow equations are invariant under simultaneous sign reversals of any two couplings, we can always choose $\lambda_1(0),\lambda_2(0)\geq 0$, while allowing either sign of $\lambda_3(0)$.

An important property of the flow is
\begin{equation}
    \frac{d}{d\ell}(\lambda_1\lambda_2\lambda_3)
    =-(\lambda_1^2\lambda_2^2
      +\lambda_2^2\lambda_3^2
      +\lambda_3^2\lambda_1^2)\leq 0~.
\end{equation}
Thus, the product $\lambda_1\lambda_2\lambda_3$
always decreases along the flow. 
For the $\lambda_1(0)\geq\lambda_2(0)\geq\lambda_3(0)\geq0$ configuration, this allows the initially repulsive
couplings to first decay and then possibly develop an attractive component.
For the  $\lambda_1(0)\geq\lambda_2(0)\geq 0\geq\lambda_3(0)$ configuration, the product is already
negative and can decrease toward negative infinity.

We further note that the differences of the squared
couplings are conserved,
\begin{equation}
    \frac{d}{d\ell}(\lambda_i^2-\lambda_j^2)=0.
\end{equation}
Defining
\begin{equation}
    \delta_{13}\equiv\lambda_1^2-\lambda_3^2\geq 0,\qquad
    \delta_{23}\equiv\lambda_2^2-\lambda_3^2\geq 0,
\end{equation}
the three coupled flow equations reduce to a single
differential equation,
\begin{equation} \label{eqn:lambda3_flow}
    \frac{d\lambda_3}{d\ell}
    =-\sqrt{(\delta_{13}+\lambda_3^2)
            (\delta_{23}+\lambda_3^2)}.
\end{equation}
Here we have used $\lambda_1,\lambda_2\geq 0$, whose signs cannot change because of the conserved
quantities. 
The exact solution is then given by
\begin{equation}
    \ell=\int_{\lambda_3(\ell)}^{\lambda_3(0)}
    \frac{dx}{
    \sqrt{(x^2+\delta_{13})(x^2+\delta_{23})}}.
    \label{eqn:lambda3_integral}
\end{equation}

Since the right-hand side of Eq.~\eqref{eqn:lambda3_flow} is strictly negative, $\lambda_3$ decreases monotonically.
For the $\lambda_3(0)>0$ configuration, it crosses zero at a finite RG scale
\begin{equation}
    \ell_*=\int_0^{\lambda_3(0)}
    \frac{dx}{
    \sqrt{(x^2+\delta_{13})(x^2+\delta_{23})}}~,
\end{equation}
which is finite for $\delta_{13},\delta_{23}>0$. 
The flow eventually diverges at
\begin{equation}
    \ell_c=\int_{-\infty}^{\lambda_3(0)}
    \frac{dx}{
    \sqrt{(x^2+\delta_{13})(x^2+\delta_{23})}}.
\end{equation}
The integral is finite for $\delta_{13},\delta_{23}>0$.
Near the divergence, $\delta_{13},\delta_{23}$ are negligible compared with $\lambda_3$, giving
\begin{equation}
    (\lambda_1,\lambda_2,\lambda_3)
    \simeq \frac{(1,1,-1)}{\ell_c-\ell}.
\end{equation}
Therefore, generic initial conditions flow toward
the same strong-coupling trajectory, characterized
by two repulsive and one attractive coupling.

For $\delta_{23}=0, \delta_{13}>0$, $l_*$ diverges and $\lambda_3$ does not change sign.  
The flow instead approaches a finite fixed point,
\begin{equation}
    (\lambda_1,\lambda_2,\lambda_3)
    \longrightarrow (\sqrt{\delta_{13}},0,0).
\end{equation}

For the fully symmetric initial condition
$\lambda_1(0)=\lambda_2(0)=\lambda_3(0)>0$, the three equations reduce to
\begin{equation}
    \frac{d \lambda}{d\ell}=-\lambda^2,\qquad
    \lambda(\ell)=\frac{\lambda(0)}{1+\lambda(0)\ell}.
\end{equation}

\subsection{Fixed rays for general $\alpha$. }

For small but finite $\alpha$, we expect the same two classes of solutions.
We consider solutions of the form
\begin{equation}
    \lambda_i(\ell)\sim\frac{\Lambda_i}{\ell_c-\ell}.
\end{equation}
Substituting into Eq.~\eqref{eqn:triflow}, we obtain two different classes of relevant solutions: 
\begin{equation} \label{eqn:rays}
\begin{aligned}
    &\text{Symmetric:}\quad
    (\Lambda_1,\Lambda_2,\Lambda_3)
    =\frac{1}{\alpha-1}(1,1,1),
    &\text{Mixed-sign:}\quad
    (\Lambda_1,\Lambda_2,\Lambda_3)
    =\frac{1}{\alpha^2+\alpha+1}
    (\alpha+1,\alpha+1,-1).~ 
\end{aligned}
\end{equation}
For $-1<\alpha<1$, the symmetric solution flows to zero for repulsive interactions but diverges at a finite $\ell_c$ for attractive interactions.
The mixed-sign solutions describe two repulsive and one attractive coupling.

We now check the stability of the two sets of solutions. 
Introducing small perturbations
$\lambda_i(\ell)= \frac{\Lambda_i}{\ell_c-\ell} +\delta\lambda_i(\ell)$, 
we obtain
\begin{equation}
    \frac{d}{d\ell}\begin{pmatrix}
    \delta\lambda_1\\\delta\lambda_2\\\delta\lambda_3 \end{pmatrix}
    =\frac{1}{\ell_c-\ell} \begin{pmatrix}
        2\alpha\Lambda_1& -\Lambda_3 & -\Lambda_2\\
        -\Lambda_3 & 2\alpha\Lambda_2& -\Lambda_1\\
        -\Lambda_2 & -\Lambda_1 & 2\alpha\Lambda_3
    \end{pmatrix}
    \begin{pmatrix}
    \delta\lambda_1\\\delta\lambda_2\\\delta\lambda_3 \end{pmatrix}~.
\end{equation}

For the symmetric ray, the perturbations decompose into a symmetric mode and two degenerate $C_3$-breaking modes, as in the main text. 
The symmetric mode satisfies
$\delta\lambda_1=\delta\lambda_2=\delta\lambda_3$
and scales as
\begin{equation}
    \frac{\delta\lambda_i}{\lambda_i}\propto(\ell_c-\ell)^{-1}.~
\end{equation}
This perturbation vanishes in the repulsive flow as $\ell\to\infty$, while it diverges as $\ell\to\ell_c$ along the attractive ray.
The latter divergence, however, simply corresponds to a shift of the critical scale $\ell_c$ and does not signal an instability of the ray.

The two $C_3$-breaking modes satisfy $\delta\lambda_1+\delta\lambda_2+\delta\lambda_3=0$, with
\begin{equation}
    \frac{\delta\lambda_i}{\lambda_i}
    \propto
    (\ell_c-\ell)^{(\alpha+2)/(1-\alpha)}.
\end{equation}
For $-1<\alpha<1$, the exponent is positive. 
The relative perturbations diverge as $\ell\to\infty$ for the repulsive flow and vanish as $\ell\to\ell_c$ along the attractive flow. 
Therefore, the fully attractive symmetric ray is stable against $C_3$-breaking perturbations, whereas the fully repulsive symmetric flow is unstable.

For the mixed-sign ray, the linearized equations
give two independent perturbation modes.
The first satisfies
$\delta\lambda_1=-\delta\lambda_2$ and
$\delta\lambda_3=0$, corresponding to an asymmetry
between the two repulsive couplings. Its relative
magnitude scales as
\begin{equation}
    \frac{\delta\lambda_1}{\lambda_1}
    \propto(\ell_c-\ell)^{
    \frac{(1-\alpha)(\alpha+2)}
    {\alpha^2+\alpha+1}}.
\end{equation}
The second mode satisfies
$\delta\lambda_1=\delta\lambda_2$ and
$\delta\lambda_3=2(\alpha+1)\delta\lambda_1$,
describing a change in the relative strengths
of the repulsive and attractive couplings.
For this mode,
\begin{equation}
    \frac{\delta\lambda_1}{\lambda_1}
    \propto(\ell_c-\ell)^{
    \frac{(\alpha+1)(\alpha+2)}
    {\alpha^2+\alpha+1}}.
\end{equation}
Both exponents are positive for $-1<\alpha<1$, so the relative perturbations vanish near the divergence. 
The mixed-sign ray is therefore stable against small deviations.

\end{document}